\documentclass[manuscript]{aastex}
\usepackage{graphicx,lscape}
\slugcomment{Submitted to AJ, December 22 2011}

\shorttitle{Nearby dwarf galaxies}
\shortauthors{McConnachie}

\begin{document}

\title{The observed properties of dwarf galaxies in and around the Local Group}

\author{Alan W. McConnachie\email{alan.mcconnachie@nrc-cnrc.gc.ca}}
\affil{NRC Herzberg Institute of Astrophysics, 5071 West Saanich Road,
  Victoria, B.C., V9E 2E7, Canada}

\clearpage
\newpage

\begin{abstract}
  Positional, structural and dynamical parameters for all dwarf
  galaxies in and around the Local Group are presented, and various
  aspects of our observational understanding of this volume-limited
  sample are discussed. Over 100 nearby galaxies that have distance
  estimates reliably placing them within 3\,Mpc of the Sun are
  identified. This distance threshold samples dwarfs in a large range
  of environments, from the satellite systems of the MW and M31, to
  the quasi-isolated dwarfs in the outer regions of the Local Group,
  to the numerous isolated galaxies that are found in its
  surroundings. It extends to, but does not include, the galaxies
  associated with the next nearest groups, such as Maffei, Sculptor,
  and IC342. Our basic knowledge of this important galactic subset and
  their resolved stellar populations will continue to improve
  dramatically over the coming years with existing and future
  observational capabilities, and they will continue to provide the
  most detailed information available on numerous aspects of dwarf
  galaxy formation and evolution. Basic observational parameters, such
  as distances, velocities, magnitudes, mean metallicities, as well as
  structural and dynamical characteristics, are collated, homogenized
  (as far as possible), and presented in tables that will be
  continually updated to provide a convenient and current on-line
  resource. As well as discussing the provenance of the tabulated
  values and possible uncertainties affecting their usage, the
  membership and spatial extent of the MW sub-group, M31 sub-group and
  the Local Group are explored. The morphological diversity of the
  entire sample and notable sub-groups is discussed, and time-scales
  are derived for the Local Group members in the context of their
  orbital/interaction histories. The scaling relations and mean
  stellar metallicity trends defined by the dwarfs are presented, and
  the origin of a possible ``floor'' in central surface brightness (and, more
  speculatively, stellar mean metallicity) at faint magnitudes is considered.
\end{abstract}


\clearpage

\newpage

\section{Introduction}

There has been a veritable explosion of data, discoveries and
realizations in the past decade relating to the broad
research area of the structure and content of nearby galaxies,
particularly those in which it is possible to resolve the individual
stars that contribute to the galaxy's luminosity, and to which this
article exclusively refers. By necessity, the objects of the attention
of this research are the Milky Way, its satellites, and their
neighbours within the so-called Local Group, but it is ever expanding
to include the far richer sample of objects that fall within the
loosely-defined part of the Universe referred to only as the Local
Volume.

The rapid growth in interest in this field can be traced to two,
distinct, developments that have occurred in tandem. The first is
entirely observational: nearby galaxies subtend large areas on the
sky, and contain a high density of individually rather-faint stellar
sources. Thus, the advent of wide field, digital, high resolution,
multiplexing capabilities on large telescopes (and the computational
facilities to handle the resulting data-flow) have had a profound
effect (and will continue to do so). The second is the result of an
increased understanding of the nature of our immediate cosmic
environment, facilitated in no small part by computational tools
(hard- and soft-ware) that have allowed for a higher-resolution
examination of the consequences of the prevailing cosmological
paradigm. The discoveries and realizations that these advances have
enabled are significant: the hierarchical nature of galaxy
formation, and the realization that some of the relics of that process
may still be identifiable at $z = 0$; the quest for a better
understanding of dark matter, and the realization that the smallest
galaxies nearest to us may be the darkest laboratories in the
Universe; the chemical evolution of the Universe, and the realization
that the nucleosynthetic imprints of the earliest generations of star
formation may still be found in the Galaxy and its satellites. These
are but a few examples. A new phrase has developed to collectively
describe the aforementioned research and its motivations, and which
contrasts it with its higher-redshift cousins: ``near-field
cosmology''.

The phrase ``near-field cosmology'' is relatively recent, and dates
from the Annual Review article of \cite{freeman2002}. As such, it is
sometimes considered a new field within astrophysics as a whole,
although this is grossly unfair to its early exponents. Indeed, what
could be argued as two of the most important papers in astrophysics --
that of \cite*{eggen1962} and \cite{searle1978} -- are quintessential
examples of observational near-field cosmology. The former took a
well-defined sample of stars for which high quality spectroscopic and
photometric data existed and from that motivated a collapse model for
the formation of the Galaxy. The latter used complexities within the
stellar populations of the Galactic globular cluster system to
motivate a scenario where the Galaxy (or part of it) is built through
the accretion of smaller stellar systems. The principle ideas from
both papers continue to form critical aspects of modern galaxy
formation theory.

Many more galaxies are now able to have their resolved stellar content
examined than was possible at the time of these early studies. The
content and structure of the Andromeda Galaxy (M31) provides an
excellent comparison (and, increasingly, contrast) to our own Galaxy
(e.g., \cite{courteau2011,huxor2011,
  watkins2010,mcconnachie2009b,yin2009,ibata2007,chapman2007,kalirai2006b,koch2008}
and references therein). Even excluding this prominent astronomical
target, however, there are still nearly a hundred or so galaxies for
which their individual stars can be brought into focus by astronomers
wanting to unearth fossil signatures of galactic formation.

The overwhelming majority of nearby galaxies -- indeed, the
overwhelming majority of galaxies in the entire Universe -- are dwarf
galaxies. Exactly how low a luminosity or how small a mass is required
before a galaxy is deemed meager enough to warrant this popular
designation is generally not well-defined or
accepted. \cite{kormendy1985} famously showed that when surface
brightness is plotted against luminosity for elliptical galaxies, two
relations are apparently formed, one for ``giant'' galaxies, and one
for ``dwarf'' galaxies. This has usually been interpreted as evidence
that these systems should be separated into two distinct groups (in
terms of different formation mechanisms). Some more recent studies, on
the other hand, suggest that there is actually a continuous sequence
connecting the dwarf and giant regimes, and that the change in the form of the scaling
relations with luminosity does not necessitate fundamental differences
between low and high luminosity systems (e.g., \citealt{graham2003}
and references therein). With these considerations in mind, and for
the purpose of this article, I consider dwarfs to have absolute visual
magnitudes fainter than $M_V \sim -18$ (also adopted by
\citealt{grebel2003}). In what follows, this places all nearby galaxies
but the Milky Way, the Large Magellanic Cloud, M31, M33, NGC55 and
NGC300 in the dwarf category.

The closest galaxies to the Milky Way are its own satellites, the
most prominent of which are the Large and Small Magellanic Clouds
(L/SMC), both of which are naked eye objects. The earliest reference
to the Magellanic Clouds that survives is by the Persian astronomer
Abd al-Rahman al-Sufi in his 964 work, {\it ``Book of Fixed Stars''}
(which also includes the earliest recorded reference to the Andromeda
Galaxy). A millennium or so later, \cite{shapley1938a} discovered ``a large rich
cluster with remarkable characteristics'' in the constellation of
Sculptor. He speculated that it was

\begin{quote}
``...a super-cluster of the globular type and of galactic dimensions;
or a symmetrical Magellanic Cloud devoid of its characteristic bright
stars, clusters, and luminous diffuse nebulosity; or a nearby
spheroidal galaxy, highly resolved, and of abnormally low surface
brightness. These phrases are merely different ways of describing the
same thing and of pointing out the uniqueness of the object.''
\end{quote}

\noindent Over seventy years later, a large body of research still
devotes itself to understanding the detailed properties, structural,
chemical and dynamical characteristics of this class of stellar
system. The prototype described above is referred to as the Sculptor
dwarf spheroidal (dSph) galaxy.

Excellent discussions of the observational properties of -- and our
astrophysical understanding of -- the nearest (Local Group) dwarf
galaxies can be found in a large number of dedicated review articles
(e.g.,
\citealt{gallagher1994,grebel1997,mateo1998a,vandenbergh2000,geisler2007,tolstoy2009}). The
compilations of observational data presented in numerous works by
S. van den Bergh, in particular \cite{vandenbergh2000}, and the Annual
Review article by \cite{mateo1998a} are particularly valuable
resources for the astronomical community. Since their original
publication, however, there have been notable advances (in part a
product of the observational and theoretical factors described
earlier). Not least of these is the {\it more-than-doubling} of the
number of known galaxies in the Local Group. In the surrounding of the
Local Group too, there have been significant new discoveries, and
techniques and observations that used to be applicable only to the
Milky Way satellites or nearby Local Group galaxies can now be applied
to these more distant relatives.

The ability to reach the resolved populations of ever more distant
galaxies is important not just for the improvement in statistics that
it offers, but also in terms of the increasing range of environments
that it allows us to study. The overwhelming amount of detailed
information regarding the stellar population properties of dwarf
galaxies comes from studies of the MW satellite systems. Given the
sensitivity of low mass systems to both internal (e.g., feedback
through star formation) and external (e.g., ram and tidal stripping)
environmental mechanisms, this might be expected to bias our
understanding towards conditions that hold only for the MW
environment. Within the very nearby universe, however, there are a
large number of environments able to be probed, from the surroundings
of the MW and M31, to the more isolated, outlying members of the Local
Group, to the very isolated dwarf galaxies that surround the Local
Group, to the numerous nearby groups that populate and form our
immediate neighbourhood. A large number of dedicated surveys in the
past few years are systematically examining galaxies in the Local
Volume at a range of wavelengths, and which include among their
targets some of the more distant galaxies in this sample. These
surveys are obtaining precision insights into local galaxies that naturally
compliment lower resolution studies of their more numerous, but more
distant, cousins (e.g., SINGS, \cite{kennicutt2003}; the Spitzer LVL
(\citealt{dale2009}; THINGS, \cite{walter2008}; FIGGS,
\cite{begum2008a}; HST ANGRRR/ANGST, \cite{dalcanton2009}; the GALEX
survey of galaxies in the Local Volume, \cite{lee2011}; the H$\alpha$
survey of \cite{kennicutt2008}; and many others).

I have alluded to several motivations for pursuing studies of nearby
galaxies, be they related to cosmic chemical processing, dynamical
evolution, galactic formation mechanisms, environment, near field
cosmology or a wealth of other research areas on which much can be
written.  It is not the purpose of this article to contribute to those
discussions here; the recent Annual Review article by
\cite{tolstoy2009} provides an excellent, current overview of the
status of research into many of these topics using Local Group
galaxies. Rather, this article attempts to collate, homogenize and
critically review our observational understanding of aspects of the
nearby populations of dwarf galaxies, with particular emphasis on the
origins and uncertainties of some key observable parameters. I will
discuss science topics only when such discussions relate directly to
observational measurements and those measurements, in my opinion,
provide grounds for caution, or when such discussion provides a
context for highlighting observational relationships between
parameters. This article primarily focuses on the stellar properties
of the galaxies (and derived properties, such as the implied dark
matter content), and cites observations conducted mostly at
ultraviolet--optical--near-infrared wavelengths, with less discussion
on their gaseous and dust content. With the intense efforts currently
underway to provide a more full characterization of these systems, I
expect some numbers provided in this article will be out-of-date by
the time it is published. As such, tables presented herein will be
continually updated to provide a convenient,
on-line library of dwarf galaxy parameters for general reference and
use by the community.

The layout of the article is as follows. \S2 describes the selection
of the sample and gives a general summary of my methodology and
reasoning in the construction of the dataset. \S3 reviews the
discovery space of the galaxies, and discusses general issues relating
to distances, velocities and the membership (or otherwise) of
larger-scale groups/sub-groups. \S4 reviews the observed photometric
and structural characteristics of the sample and discusses aspects of
the scaling relations that they define. \S5 reviews their masses
(stellar, HI and dynamical) and internal kinematics, and discusses
issues related to morphological classifications and our dynamical
understanding of the sample. \S6 presents a compilation of available
mean stellar metallicities and discusses associated observational trends. \S7
concludes and summarizes.

\section{Constructing the dataset}

The sample of galaxies discussed in this article consist of all known
galaxies with distances determined from measurements of resolved
stellar populations (usually Cepheids, RR Lyrae, tip of the red giant
branch, TRGB, but also including horizontal branch level and
main-sequence fitting; see \citealt{tammann2008} for a thorough
discussion of the former three methods and
references) that place them within 3\,Mpc of the Sun.  Consequently,
this sample contains the satellite systems of two major galaxies (the
MW and M31), the quasi-isolated, outlying members of the Local Group,
and the nearby isolated dwarf galaxies that do not clearly belong to
any major galaxy grouping. It is therefore expected that this sample
may be valuable (as, indeed, various individual members and sub-sets
have already proven valuable) for examining the role played by
environment in dwarf galaxy evolution, as discussed in \S1. There is
also a practical motivation for the choice of distance limit; namely,
that a much larger threshold will start to select members of nearby
groups (the closest of which is the Maffei group at $\sim 3$\,Mpc,
although its awkward location at low Galactic latitudes prevents
accurate distances being determined for its members; see
\citealt{fingerhut2007}), and would expand the scope of this article
considerably.

Of course, there are no solid boundaries between galaxy groups,
filaments and the field, and some of the galaxies discussed herein are
likely dynamically associated with neighboring structures like the
Sculptor (e.g., \citealt{karachentsev2003a}) or IC342 (e.g.,
\citealt{karachentsev2003b}) groups, for example. Nevertheless, at the
time of writing, the adoption of these selection criteria identifies
exactly 100 galaxies (not including the MW and M31), of which 73 are
definite or very likely members of the Local Group. It is likely
that our basic knowledge of all of these galaxies and their resolved
stellar content can be improved
dramatically over the coming years with existing observational
capabilities, and the more distant galaxies provide stepping stones
into the Local Volume for future resolved stellar population studies
enabled by next generation facilities like the thirty-meter class
telescopes. It is with an eye to these latter considerations that the
distance is not limited to the zero-velocity surface of the Local
Group ($R_{LG}$).

The ensuing discussion will focus on dwarf galaxies. Of the 102
galaxies that are listed in the subsequent tables, virtually no
discussion will be given to the Milky Way and M31, and numbers
relating to M33, NGC55 and NGC300 are included for completeness
only. The same caveat applies to the Magellanic Clouds, since here
the research body and available data are so extensive and far exceeds that
which exists for most other dwarf-like systems that the reader is referred
to the many review articles, books, and conferences that deal
specifically with these bodies (e.g.,
\cite{westerlund1997,vanloon2009} and references therein).

In compiling this catalog of galaxies, and in addition to the papers
cited in each table, I have made extensive use of the NASA
Extragalactic Database (NED), the HyperLeda database
(\citealt{paturel2003}), \cite{mateo1998a}, \cite{vandenbergh2000},
and \cite{karachentsev2004}. The requirement that all galaxies have
distances based on resolved stellar populations leads to the exclusion
of a few galaxies that have low velocities that could potentially
place them within 3\,Mpc, but which lack any direct distance
measurement (e.g., see Table~1 of \citealt{karachentsev2004}). The low
latitude galaxy Dwingeloo~1 is also excluded by this criterion, since
its distance of $\sim 2.8$\,Mpc is based on a Tully-Fisher estimate
(\citealt{karachentsev2003b}; this galaxy is very likely a member of
the Maffei group). The (uncertain) TRGB measurement of MCG9-20-1 by
\cite{dalcanton2009} places it at a distance of $\sim 1.6\,$Mpc,
although its radial velocity of $v_\odot \sim 954$\,km\,s$^{-1}$ is
very high if it is really this close, and the photometric data now
seem to favor a larger distance (K. Gilbert \& J. Dalcanton, private
communication), so it too has been excluded from the list. Finally,
there are a few galaxies for which it is unclear if they lie closer or
further than 3\,Mpc (e.g., UGCA 92, $D = 3.01 \pm 0.24$\,Mpc;
\citealt{karachentsev2006}). I exclude such galaxies, even if the
quoted uncertainties would bring them within range of the distance
cut. Of course, if future estimates should clearly indicate a closer
distance, then the on-line version of the tables will be updated as
appropriate.

In compiling these references, I have favored, where possible, papers
that are based on analysis of the resolved stellar populations. In the
interest of homogeneity and in an (ill-fated) attempt to minimize the
inevitable systematic differences between measurements for different
galaxies, I also generally try to limit the number of distinct
publications, studies and/or methodologies that contribute to the
overall dataset, favoring large surveys of multiple galaxies in
preference to studies of individual galaxies. However, if there appear
to be significant discrepancies in the literature over observed values,
then these differences are indicated in either the footnotes and/or
the text. Infrequently, I have estimated uncertainties
where none were given in the original publications.

My primary intent in compiling this dataset is to provide useful
information on the population of (dwarf) galaxies in the nearby
Universe. I have tried to ensure that the references collected herein
-- while not complete in any way -- are generally recent enough and
relevant enough that I hope they will be able to provide the reader
with a starting point from which most of the germane literature can be
traced.

\section{Basic properties, positions and velocities}

Table~1 lists basic information for all nearby galaxies that satisfy the selection criteria
discussed in the previous section:

\begin{itemize}
\item Column 1: Galaxy name;
\item Column 2: Common alternative names;
\item Column 3: Indicator whether they are associated with MW [G], M31
  [A], the Local Group [L], or are nearby neighbors [N];
\item Column 4: Morphological (Hubble) type. The distinction between
  dwarf elliptical (dE) and dwarf spheroidal (dSph) is based solely on
  luminosity, and is therefore somewhat arbitrary. Objects fainter
  than $M_V \sim -18$ are given a ``d'' (dwarf) prefix, and is again
  somewhat arbitrary;
\item Column 5, 6: Celestial coordinates (J2000);
\item Column 7: Original publication. For more details relating to the
  discoveries of the NGC/IC galaxies, the reader is referred to
  \cite{steinicke2010}\footnote{See also
    http://www.klima-luft.de/steinicke/index\_e.htm};
\item Column 8: Comments.
\end{itemize}

Table~2 lists position and velocity information for the galaxy sample:

\begin{itemize}
\item Column 1: Galaxy name;
\item Columns 2, 3: Galactic coordinates $(l, b)$;
\item Column 4: Foreground extinction, E(B-V): These correspond to
  values derived by \cite{schlegel1998} from an all-sky COBE/DIRBE and
  IRAS/ISSA 100\,$\mu$m map, at the coordinates of the centroid of
  each galaxy. While \cite{schlegel1998} estimate that values for
  E(B-V) are generally accurate to 16\%, a major (and sometimes
  overlooked, including by this author!) caveat for their use in the
  direction of nearby (large) galaxies is that the brightest galaxies
  at IRAS 60\,$\mu$m\footnote{specifically, the 70 galaxies listed in
    Table~2 of \cite{rice1988} with fluxes greater than or equal to
    0.6\,Jy} were excised from these maps and replaced by the ``most
  likely value'' of the underlying 100\,$\mu$m emission (see \S4.2 of
  \citealt{schlegel1998}). The resulting values are not, therefore,
  directly observed, and any variation across the area occupied by the
  galaxy is not traced by these maps. For this compilation, the
  affected galaxies are NGC6822, IC1613, M33, NGC205, NGC3109, NGC55
  and NGC300. The LMC, SMC and M31 (and by necessity also M32) were
  not excised;
\item Columns 5, 6: Distance modulus, heliocentric distance: All
  distance moduli are based on resolved stellar population analysis
  (generally Cepheid, RR~Lyrae or TRGB
  measurements, but also horizontal branch level and main-sequence
  fitting). Note that many TRGB estimates do not include the
  formal uncertainty in the absolute magnitude of the TRGB (\citealt{bellazzini2001,
    bellazzini2004a, rizzi2007});
\item Column 7: Heliocentric velocity: For most Local Group galaxies,
  the velocity corresponds to the optical velocity of the galaxy. For
  some of the more distant galaxies, velocities are measured from HI;
\item Column 8 -- 10: Distance-velocity pairs for the Galactocentric
  (MW), M31 and Local Group frames-of-reference: These are calculated
  by first assuming that the barycenter of the Local Group is located
  at the mid-point of the vector connecting the MW and M31 (for an
  adopted M31 distance modulus of $(m-M)_0 = 24.47$;
  \citealt{mcconnachie2005a}). The latter galaxy is more luminous
  (e.g., \citealt{hammer2007} and references therein), but dynamical
  mass estimates for both of these galaxies have produced conflicting
  results regarding which is more massive (e.g.,
  \citealt{little1987,kochanek1996,wilkinson1999,evans2000a,evans2000b,klypin2002,ibata2004,geehan2005,watkins2010,mcmillan2011}),
  a result nearly exclusively due to the lack of a sufficient number
  of dynamical tracers at large radii. Given this uncertainty, I
  conclude that assuming their masses to be roughly the same is
  reasonable, simple and convenient. Velocities are converted to the
  Local Group frame using the Solar Apex derived by
  \cite{karachentsev1996} and subtracting the component of the MW's
  velocity in the direction of each dwarf. M31-centric velocities are
  obtained by subtracting the component of the Local Group-centric
  velocity of M31 in the direction of each dwarf. Galactocentric
  distances and velocities are calculated in the usual way, for an
  adopted Galactic rotation velocity at the Sun of 220\,km\,s$^{-1}$
  at a radius of 8.5kpc from the Galactic center;
\item Column 11: References.
\end{itemize}

Figure~1 shows Aitoff projections of the Galactic coordinates of all
galaxies in the sample. The top panel shows only those objects
identified as definite or likely MW satellites; the middle panel shows
definite or likely M31 sub-group members (blue points) and quasi-isolated
Local Group galaxies (green points); the bottom panel shows galaxies
surrounding the Local Group within a distance of 3\,Mpc. Some
confirmed members of all galaxy groups within 5\,Mpc are also shown in
the bottom panel (grey points) to highlight the location of the Local
Group and its neighbors with respect to these nearby structures
(\citealt{karachentsev2005}). In this respect, it is important to
highlight the work of I. Karachentsev, B. Tully and their colleagues
in both the identification of nearby galaxies and groups, and in the
determination of accurate distances to many of our neighbours in order
to understand the structure of the Local Volume (e.g., see
\citealt{karachentsev1994,karachentsev2003a,karachentsev2004, tully1987,tully2006,tully2009}).

\subsection{Membership of the MW sub-group}

Figure~2 shows Galactocentric velocity versus distance for all
galaxies within 600\,kpc for which necessary data exist. Here, the
Galactocentric velocity listed in Table~2 has been multiplied by a
factor of $\sqrt3$ to provide a crude estimate of the unknown
tangential velocities of these galaxies. Dashed curves show the escape
velocity from a $10^{12}\,M_\odot$ point mass (a reasonable
approximation to the escape velocity curve of the MW for the purposes
of this discussion). The vertical dashed line indicates the
approximate cosmological virial radius of the MW, $R_{vir} \sim
300$\,kpc (\citealt{klypin2002}; these authors define $R_{vir}$ as the radius within
which the mean density of the model dark matter halo is a factor $\Delta
\simeq 340$ larger than the critical density of the Universe).

With the obvious caveat regarding the generally unknown tangential
components of the velocities of the galaxies, Figure~2 can be used to
define membership of the MW subgroup. In particular, all galaxies
within 300\,kpc of the MW are likely bound satellites. This is with
the exception of Leo~I, which may be unbound unless its tangential
components are significantly overestimated (also see
\citealt{mateo2008,sohn2007,zaritsky1989} and references
therein). Leo~T, NGC\,6822 and Phoenix may all be bound in this simple
model, but their large distance from the MW makes their membership of
the MW subgroup ambiguous. I note that no potential satellite of the
MW or M31 (\S3.2) is clearly unbound as a result of the magnitude of
its radial velocity alone; all the potential satellites with large
radial velocities relative to their host may be bound if their
tangential velocities are sufficiently small. For the purpose of the
classification in Column 3 of Table~1, and in the forthcoming
discussion, I only include as definite MW satellites those objects
that are bound by the dashed curves in Figure 2 and which have $D_G
\le R_{vir}$. It is
interesting to note that the radial distribution of satellites is such
that they are found at all radii out to $\sim 280$\,kpc from the MW,
at which point there is a notable gap in the distribution until the
next set of objects at $> 400$\,kpc. A similar gap is present at a
similar radius for satellites around M31 (\S3.2; but see the recent
discovery of Andromeda~XXVIII by \citealt{slater2011}). This tentatively
suggests that this gap could be used to observationally define the
limiting radius of the sub-groups (or equivalently, the extent of the
two host galaxies), and appears consistent with previous work and
other galaxy groups (e.g., Cen A, Sculptor, IC342:
\citealt{karachentsev2002b,karachentsev2003a,karachentsev2003b}; see
also \citealt{grebel2003}).

Pisces II is the most recently discovered satellite to the MW
(\citealt{belokurov2010}), and does not appear in Figure~2 since it
currently lacks a velocity measurement. At the time of writing, 15 new
satellites have been discovered since 2005, all of them using the
Sloan Digital Sky Survey (SDSS) and hence confined to its
footprint. \cite{irwin1994} originally argued that the MW sub-group
was essentially complete at Galactic latitudes away from the disk for
objects of comparable luminosity (or brighter) than Sextans. The fact
that only one of the new satellites (Canes Venatici) is comparable in
luminosity to any of the previously known MW satellites backs up this
claim. Indeed, \cite{koposov2008} demonstrates that most of the newly
discovered galaxies are at the limits of detection of SDSS, and
implies that -- even within the SDSS footprint -- a large number of
galaxies still await discovery. This was explored further by
\cite{tollerud2008}, who conclude that there may by many hundreds
of unseen faint satellites.

It is worth emphasizing the problems caused by the Galactic disk in
searching for satellites. Figure~3 shows the distribution of MW
satellites with (absolute) latitude. The histogram shows the observed
distribution of all 27 satellites, whereas the dashed line indicates a
uniform distribution, normalised to match observations at
$|b|>30^\circ$. There are 3 satellite at $|b| < 30^\circ$ (one of
which is Canis Major, on which there is still considerable debate as
to whether it is a dwarf galaxy or an unrelated disk substructure;
e.g., \citealt{martin2004a,martin2004b,momany2004,martinezdelgado2005,moitinho2006,butler2007,lopezcorredoira2007}), a number that is very
low considering this accounts for half of the sky. Obscuration by the
Galactic disk, however, does not become a serious problem until
$|b|\lesssim20^\circ$. While some of the area probed by SDSS reaches
to very low Galactic latitudes, the majority of the Legacy area is
located at $b>30^\circ$, likely explaining the dearth of satellites at
intermediate latitudes.

Two photometric surveys currently underway -- the Pan-STARRS
(\citealt{kaiser2002}) PS1 survey in the North and SkyMapper
(\citealt{keller2007}) in the South -- together will map the entire
sky to equivalent or deeper magnitude limits to SDSS, and consequently
should reveal the presence of many more satellites. These should
remove the spatial bias introduced by SDSS. In addition, other
wide-field (but not all-hemisphere) surveys such as that made possible
with the soon-to-be-commissioned Subaru/HyperSuprimeCam will allow for
even fainter satellites to be detected (if they exist). In the more
distant future, the Large Synoptic Survey Telescope (LSST) will
re-examine the southern hemisphere sky to even greater depth. Judging
by recent discoveries, there is every reason to expect large numbers
of very faint satellites will be discovered by all these
projects. However, objects behind the Galactic plane will likely
remain hidden, and it is unlikely that (m)any bright ($M_V<-8$)
Galactic dwarf galaxies at higher Galactic latitudes remain to be
discovered

\subsection{Membership of the M31 sub-group}

Figure~4 shows M31-centric velocity versus distance for all galaxies
within 600\,kpc of Andromeda for which these measurements
exist. Again, the M31-centric velocity listed in Table~2 has been
multiplied by a factor of $\sqrt3$ to account for the unknown
tangential velocities of these galaxies. Following Figure~2, dashed
curves show the escape velocity from a $10^{12}\,M_\odot$ point mass,
and the vertical dashed line indicates the approximate cosmological
virial radius of the M31, $R_{vir} \sim 300$\,kpc (e.g.,
\citealt{klypin2002}).

Whereas only Leo~I around the MW stands out as a possible unbound
satellite from its radial velocity (\citealt{mateo2008}), M31 has at
least two close-in dwarf galaxies which are likely unbound, Andromeda
XII (\citealt{chapman2007}) and XIV (\citealt{majewski2007}). The
Pegasus dwarf irregular galaxy (DIG) and IC1613 are both extremely
distant from Andromeda, but could arguably be dynamically
associated. In this respect they are similar to NGC6822, LeoT and
Phoenix around the MW. It is often overlooked that IC10 and LGS3 are
clear M31 satellites (as close to M31 as the Andromeda VI and XVI
dSphs). However, these two galaxies differ from the majority of
satellites insofar as they both have non-negligible gas fractions and
younger stellar populations. LGS3 is generally classified as being
morphologically akin to LeoT (a ``transition-type'' dwarf; see \S5 for
discussion), whereas IC10 is generally classified as a low mass dwarf
irregular. Indeed, as a whole, the M31 satellite system contrasts with
the MW satellites in the variety of morphological types present.
Around the MW, only the LMC and SMC are not classified as
dSphs. However, in addition to the dSph satellites, M31 also has a
transition dwarf (LGS3), a dwarf irregular (IC10, although it is much
lower luminosity than either of the Magellanic Clouds), three dwarf
elliptical galaxies (NGC205, 147 and 185), a compact elliptical (M32)
and a low mass spiral galaxy (M33).

The last seven years has seen a rapid growth in the number of known M31
satellites through focused wide-field studies of the environment
around Andromeda. All the new dwarfs have been discovered as
overdensities of red giant branch (RGB) stars. This, therefore,
imposes a basic magnitude limit on the galaxies that can be found,
since it requires galaxies to have enough stars that a reasonable
number are passing though the RGB phase. An exact study of the
selection effects for dwarf galaxy searches around M31 has yet to be
published, but the practical limit of these searches appears to be
around $M_V \sim -6$.  The most extensive survey of M31's environs, by
the Pan-Andromeda Archaeological Survey (PAndAS;
\citealt{mcconnachie2009b}) provides complete spatial coverage out to
a maximum projected radius from M31 of 150\,kpc. As shown in
\cite{mcconnachie2009b} and \cite{richardson2011}, the projected
radial distribution of satellites shows no sign of declining within
this radius. It is highly probable, therefore, that a large number of
satellites remain to be discovered at large projected radius (recently verified
by the discoveries of Andromeda XXVIII (\citealt{slater2011}) and
Andromeda XXIX
(\citealt{bell2011}) in the SDSS DR8). At small radius ($\lesssim
60$\,kpc), there is also a deficit of galaxies due to the dominance
of M31 stellar populations in this region making it difficult to
identify any low luminosity satellites that may be superimposed in
front of (or behind) the main body of M31.

Within both the MW and the M31 sub-group there are at least three
potential dynamical associations of satellites
(``sub-sub-groups''). Around the MW, Leo~IV and V are in close
proximity to each other (positions and velocities) and may well
constitute a binary system (\citealt{belokurov2008,dejong2010}). Within the M31 sub-group, Andromeda~XXII
is only about 40\,kpc in projection from M33, whereas it is over
200\,kpc in projection from M31 (\citealt{martin2009}). Thus Andromeda~XXII may actually be
the only known satellite of M33 (the bright dSph Andromeda II is also
considerably closer to M33 than M31, but in a region where the
gravitational potential of M31 clearly dominates). A second possible
pairing is found to the North of M31, and includes
NGC185 and 147. 
\cite{vandenbergh1998} postulated
that NGC185 and 147 were a binary system; they are separated by only 1
degree in projection and their distances imply a three dimensional
separation of $\sim 60$\,kpc. Their velocities differ by only
$\sim 10$\,km\,s$^{-1}$, and such a small separation in phase-space seems
unlikely (but not impossible) if these galaxies are not, or never have
been, a binary system. 

\subsection{Membership of the Local Group, and identification of
  nearby neighbors}

Figure~5 shows Local Group-centric velocity versus distance from the
barycenter of the Local Group for all galaxies in the sample for which
these data exist. The Local Group-centric velocity listed in Table~2
has been multiplied by a factor of $\sqrt3$ to account for the unknown
tangential components. Dashed curves indicate the escape velocity from
a point mass of $2 \times 10^{12}\,M_\odot$. Dotted curves indicate
the escape velocity from a point mass of $5 \times 10^{12}\,M_\odot$,
the approximate total dynamical mass of the Local Group as implied from the timing
argument (e.g., \citealt{lyndenbell1981}). Many of the points at
$D_{LG}\lesssim500$\,kpc represent galaxies that are satellites of
either the MW or M31. These systems are dynamically dominated by the gravitational
potential of these massive hosts rather than the Local Group as a
whole. 

There is a clear separation in Figure~5 between outer, apparently
bound, Local Group members (the outermost of which is UGC 4879), and
the next nearest galaxies (the four members of the NGC3109 grouping,
see \citealt{vandenbergh1999b}). The rise in velocity with distance
for all the nearby neighbors is due to the Hubble expansion, and the
reader is referred to \cite{karachentsev2009} for a beautiful analysis
of the Hubble flow around the Local Group.  

The six outermost Local Group members cluster around $v_{LG} = 0$, and
their mean velocity is consistent with zero ($<v> = 4 \pm
16$\,km\,s$^{-1}$, where the uncertainty is the random error in the
mean). As such, their mean distance can be used as an estimate of the
zero-velocity surface of the Local Group, and is $R_{LG} = 1060 \pm
70$\,kpc (where the uncertainty is the random error in the mean;
cf. \cite{courteau1999,karachentsev2002a,karachentsev2009} and
others). It is also worth pointing out that the velocity dispersion of
the Local Group can be estimated using those Local Group galaxies that
are far from either the MW or M31 and that have published
velocities. The six Local Group galaxies identified in Figure 5, with
the addition of IC1613 and Cetus, are all more than 500\,kpc from
either of the two large galaxies, and yield $\sigma_{LG} = 49 \pm
36$\,km\,s$^{-1}$ (where the uncertainty is the random error in the
dispersion). Note that there is a systematic error component in all
these estimates caused by our choice for the location of the
barycenter of the Local Group, although it has a generally weak effect
on any values given here. The fact that the measured velocity
dispersion for the entire Local Group is so low that it is of order
the velocity dispersion of the gas in the SMC is peculiar and well
noted (e.g., \citealt{sandage1986} and references therein). It may
result from the necessity to measure the dispersion from galaxies that
are mostly at or near turn-around.

Not including satellites of the MW and M31, it is highly likely that
the census of Local Group galaxies remains severely incomplete. Unlike
recently discovered MW satellites, isolated Local Group galaxies are
not nearby. Unlike the M31 sub-group, isolated Local Group galaxies do
not cluster in a specific area of sky amenable to dedicated
surveys. The middle panel of Figure~1 shows that 8 of the 13
(quasi-)isolated Local Group galaxies are all in one quadrant of the
sky (the same one that contains M31). However, it is also not clear
what the expected spatial distribution of non-satellite Local Group
galaxies should be, and it is likely naive to expect an isotropic
distribution. Heroic searches for new Local Group galaxies and
nearby neighbours have been made by groups lead by I. Karachentsev,
V.  Karachentseva (e.g., \citealt{karachentseva1998,karachentseva1999,karachentsev2000}
and many others) and A. Whiting (\citealt{whiting1997,whiting1999})
through the visual inspection of
literally every plate of the second Palomar Observatory Sky Survey and
the ESO/Science Research Council survey, and these remain the most
comprehensive all-sky searches to-date. \cite{whiting2007} attempt to
quantify their visual search of these plates; they estimate that their
search is $\sim80\%$ complete in the region away from interference
from the MW disk (more than 70\% of the sky) to a surface brightness
level approaching 26\,mags\,arcsec$^{-2}$. This translates to (at
most) 1 or 2 currently uncatalogued Local Group members down to this
limit, away from the disk of the Galaxy. Current and future
large-scale digital surveys of the sky - including blind HI surveys as
well as optical studies - are potentially rich hunting
grounds to try to improve the census of nearby galaxies, be they
members of the Local Group or the nearby neighbors that help define
its environment.

\section{Photometric and structural parameters}

Table~3 presents global photometric and structural parameters for each galaxy
in the sample:

\begin{itemize}
\item Column 1: Galaxy name;
\item Column 2: Apparent magnitude in V (Vega magnitudes). In general,
  these have been corrected for foreground extinction (but not
  internal extinction), although in some cases
  the original papers are unclear as to whether extinction corrections
  were applied to the apparent magnitudes. For many of the fainter
  galaxies that have been resolved into stars, the magnitude may not
  be based on integrated light (often undetectable) and is instead estimated by measuring
  the luminosity of the brighter stars and by making assumptions
  regarding the luminosity function of stars below the detection
  limit;
\item Column 3: Radius (arcmins) containing half the light of the
  galaxy, measured on the semi-major axis. A vast array of different
  scale-radii are used in the literature, and in cases where the original
  papers use other scale-radii (e.g., core, exponential,
  half-brightness, $R_{25}$, etc.), I have derived half-light radii by
  integrating under the appropriate profile, normalised to the
  apparent magnitude (and with the appropriate ellipticity if
  measured, else assumed to be circular). In
  cases where the original papers did not fit a profile, I have assumed
  an exponential profile. These latter estimates are particularly
  uncertain and in these cases I have also given the original scale
  radius in Column~12;
\item Column 4: Central surface brightness
  (mags\,arcsec$^{-2}$). Where possible, I cite directly measured
  values, although in many cases (including in the original papers)
  this parameter is derived by normalizing the measured profile to the
  apparent magnitude;
\item Column 5: Position angle of major axis, measured in degrees east
  from north. For systems that show a change in this quantity as a
  function of radius, a mean value has been estimated;
\item Column 6: Ellipticity $\epsilon = 1 - b/a$, where $b$ is the
  semi-minor axis and $a$ is the semi-major axis. For systems that
  show a change in this quantity as a function of radius, a mean value
  has been estimated;
\item Column 7: Absolute visual magnitude, derived from Column 2 by subtraction of the
  distance modulus  given in Table~2;
\item Column 8: Half-light radius in parsecs, derived from Column 3 assuming the distance
  modulus given in Table~2;
\item Column 9: The mean surface brightness within the isophote
  defined by the half-light
  radius, derived using values from Columns 2, 3 and 6. In cases where
  no ellipticity is measured, I have assumed circular isophotes;
\item Column 10: References;
\item Column 11: Comments.
\end{itemize}

For each galaxy I have tried to minimize the number of different
papers that I cite to provide complete information, with the result
that Table~3 references 45 papers from the last 21 years.  Concern
must be given to the systematic uncertainties that exist
between measurements: they are necessarily complex, they are
essentially unquantifiable, and they probably present the primary
limitation for examination of the global properties of this
population.

Despite the usefulness of parameters such as total luminosity and
half-light radius, it should be strongly stressed that these numbers
are wholly inadequate in reflecting the known complexity of galaxy
structures. It is standard practice to decompose bright galaxies into
components such as nucleus and envelope, or a bulge, a disk (several
disks?), and a halo. More recently, however, a vast number of
independent studies of galaxies included in Table~3 have shown that a
full description of their structure requires decomposition of their
surface brightness profile into multiple components (e.g.,
\citealt{irwin1995,martinezdelgado1999,lee1999b,lee1999c,lee2000,vansevicius2004,mcconnachie2007a,hidalgo2008},
among many others). The addition of spectroscopic data (see \S5) adds
new complexity to the discussion as well, since dynamically distinct
(e.g., \citealt{kleyna2003,kleyna2004}) and chemo-dynamically distinct
(e.g., \citealt{tolstoy2004,battaglia2006,battaglia2008,
  battaglia2011}) populations of stars have been shown to reside
within individual dwarf galaxies that do not necessarily always reveal
themselves as features in the global surface brightness
profiles. Thus, not only are integrated photometric properties such as
$r_h$ and $M_V$ inadequate to describe the complex structures
exhibited by many galaxies, but so too are single velocity dispersion
or rotation measurements (\S5; true for HI as well as stars; e.g.,
\citealt{lo1993,young1996,young1997a,young1997b,young2003}).

Figure 6 and 7 show basic scaling relations using the galaxy
parameters given in Table~3, first studied in detail in
\cite{kormendy1985}. In Figure 6, I plot absolute magnitude against
half-light radius. Here, I have converted the half-light radius given
in Table~3 to the geometric mean half-light radius of the major and
minor axes ($r = \sqrt{r_ar_b}$, where $r_a$ and $r_b$ are radii
measured on the major and minor axes, respectively) to account for the
presence of highly elliptical systems. The top panel includes as small
dots the location of the Galactic globular clusters, from the data
compiled by \cite{harris1996}, and the dashed line shows the direction
defined by points of constant surface brightness (averaged within the
half-light radius). I have labeled obvious outlying points, and I have
also labeled some of the least luminous (candidate) dwarf
galaxies. The lower panel shows an expansion of the region $-17 \le
M_V\le -7$.

The observed relationship between absolute magnitude -- half-light
radius shown in the top panel of Figure~6 spans a factor of one
million in luminosity. The galaxies
seem well separated from the (Galactic) globular clusters in this two
dimensional projection, although the limited number of low luminosity
galaxies (and the fact that all of them are MW satellites) complicates
interpretation of this region of the diagram. When $M_V$ and $r_h$ are
considered separately, the one-dimensional distributions do overlap,
such that the smallest, least-luminous (candidate) dwarf galaxies are as
faint as any known globular clusters and smaller than the most
extended clusters. The distinction between these stellar systems in
this parameter space becomes less clear when other types of compact
stellar systems, such as ultra-compact dwarfs and the nuclei of
early-type galaxies, are included (e.g., see Figures~5 and~8 of
\citealt{brodie2011}).

Three galaxies are clear outliers in the top panel of Figure~6: Andromeda~XIX is
extremely extended for its luminosity (see also Figure~7), whereas the
compact elliptical M32 is a known outlier in this projection due to
its extreme concentration (see also \citealt{kormendy2012} and
references therein). Interestingly, the recently identified
Local Group galaxy UGC4879 (\citealt{kopylov2008}) tends to the same
side of the relation as M32, although not as
extreme. \cite{bellazzini2011} note that UGC4879 has an unusual
structure, insofar as there appears to be two flattened wings
emanating from the central spheroid of stars, perhaps indicating the
presence of a disk. More work is needed to determine the significance
of the structure of UGC4879 and its position in this diagram.

The lower panel of Figure~6 enlarges the region $-17\le M_V\le-7$ to
demonstrate the very large scatter that exists in this regime (see
below for discussion of the fainter systems). \cite{mcconnachie2006b}
originally pointed out that, at a given luminosity, the M31 dwarf
satellites were generally larger than the MW population. With a large
number of new discoveries since this time, \cite{brasseur2011b}
demonstrate that the mean relations of the M31 and MW populations are
statistically consistent with each other. The limited number of points
contributing to the relations for these different sub-groups makes
interpretation difficult; for example, four out of ten MW satellites
have $r_h>250$\,pc, whereas 20 out of 26 M31 satellites are this
large. It may be that local environment contributes to determining the
sizes of dwarf galaxies, but perhaps the most important conclusion
that can be drawn from the lower panel of Figure~6 is the {\it lack}
of a strong relation between $M_V - r_h$ in this regime; at any given
luminosity, galaxy sizes vary by nearly an order of magnitude, and
there is only a weak dependence on luminosity.

At even fainter magnitudes, the line of constant surface brightness
(averaged within the half-light radius) included in the top panel of
Figure 6 runs approximately parallel to the relation defined by the
galaxy locus for $M_V \gtrsim -8$ (and this also has a different slope
than the weak relation at brighter magnitudes). This indicates that
the smallest galaxies so far discovered all have similar surface
brightnesses and further suggests that selection effects may be
driving the form of the $M_V - r_h$ relation at faint magnitudes. To
explore this further, I plot surface brightness against absolute
magnitude in Figure 7. The top panel shows the central surface
brightness, whereas the bottom panel uses the average surface
brightness within the half-light radius. The former is in principle
more susceptible to localized features (e.g., nuclei, individual
bright stars, etc.) than the latter. I have labeled points that are
obvious outliers in {\it both} panels (note that Bootes III is not
present in the lower panel of Figure 7 due to the lack of a measured
value for $r_h$). These galaxies are all known or suspected to be
undergoing tidal disruption.

There is a clear correlation between absolute magnitude and
surface brightness in both panels of Figure 7 for galaxies more
luminous than $M_V \sim -9$. Of considerable interest, however, is
the change in the observed relationship between surface brightness and
absolute magnitude {\it within} the dwarf regime. In particular, dwarf
galaxies fainter than $M_V \simeq -8.5$ do not continue to decrease in
surface brightness with decreasing luminosity. Instead, the surface
brightness levels off at faint magnitudes, as implied in Figure~6. The
central surface brightness distribution for the 26 galaxies fainter
than $M_V = -8.5$ in Figure~7 can be characterized by a median surface brightness
of $27.35$\,mags\,arcsec$^{-2}$ and an inter-quartile range of 1.2
magnitudes. Thus, the central surface brightness for galaxies fainter
than $M_V = -8.5$ appears remarkably constant given that the
luminosity changes by a factor of $\sim 600$ within the sample.

Surface brightness selection effects must be seriously considered in
examining the correlations in Figure~7 (see, for example, earlier
discussion by \citealt{impey1997} and references therein). Here, the
work of \cite{koposov2008} provides the most relevant reference. The
selection functions are complex and depend on distance, magnitude and
surface brightness. However, Table~3 of \cite{koposov2008} indicates
that galaxies as faint as $M_V \sim -4.4$, at distances of $\sim
180$kpc could be recovered even if their central surface brightness is as
faint as 29.9\,mags\,arcsec$^{-2}$, more than 2 magnitudes fainter than
the median observed surface brightness at faint luminosities. There is
also tentative evidence that the M31 satellites appear to behave in a
similar way to the MW satellites (although they do not extend to the
same low luminosity limits). Searches for M31 satellites are subject to
different (and as yet unquantified) selection effects compared to the
MW satellites. Taken together, it appears possible that the ``break'' in the scaling
relations at around $M_V \sim -9$ may be real. If so, it may imply
that -- as a result of either formation (e.g., inflow of gas into dark
matter haloes) or subsequent feedback/evolution (e.g., star formation)
-- there is a low-density limit to the central stellar densities of
dwarf galaxies. If the effect is instead due to selection, then we are
only observing those low luminosity galaxies with the highest surface
brightness, indicating that there may be very significant scatter in this
property at faint magnitudes. Regardless of this speculation, the
behavior of surface brightness with luminosity shown in Figure 7
certainly requires further consideration and explanation.

\section{Masses and kinematics}

Table~4 lists various information regarding the masses and kinematics
of the galaxy sample:

\begin{itemize}

\item Column 1: Galaxy name;

\item Column 2: Mass of the galaxy in stellar masses, assuming a
  stellar mass-to-light ratio of 1. This has been adopted for
  simplicity and to allow easy scaling to whatever mass-to-light ratio
  is preferred (in particular, by adopting values estimated through
  near-infrared observations);

\item Column 3: The observed velocity dispersion of the stellar
  component and its uncertainty, generally based on multiple velocity
  measurements of individual (giant) stars, else commented on in
  Column 10. Note that a single value for the velocity dispersion is
  often insufficient to describe the dynamics of the galaxy (e.g.,
  multiple components and/or radial trends);

\item Column 4: The observed rotational velocity of the stellar
  component and its uncertainty. Where available, I quote the peak
  (observed) rotational velocity, else the maximum observed rotational
  velocity\footnote{For observations that do not extend to
    sufficiently large galactocentric radius, the maximum observed rotational
    velocity will not necessarily equal the peak rotational velocity
    of the galaxy}.  In general, I try to give the rotational velocity
  uncorrected for inclination or asymmetric drift, although it is
  sometimes unclear in the original papers if these corrections have
  been applied. ``N/A'' indicates that a rotational signature has been
  looked for and none has been observed. Where only weak gradients in
  velocity are detected and/or the rotational signature is ambiguous, I
  comment on this in Column 10. In most cases, it is probably safest
  to interpret the absence of a rotational signature as the absence of
  any rotation with a magnitude equal to or greater than the magnitude
  of the velocity dispersion;

\item Column 5: Mass of HI in each galaxy. Here, I have scaled the
  values cited in the relevant papers to the distance given in Column
  6 of Table~2. Where only an upper limit from a non-detection is
  available, and that upper limit is small in comparison to the
  stellar mass, I give the HI mass as zero;

\item Column 6: The observed velocity dispersion of the HI component
  and its uncertainty. Note that the HI velocity dispersion is
  affected by local processes, such as heating due to star formation,
  and so caution is urged in interpreting these numbers. Note also
  that many galaxies are observed to contain multiple HI components (e.g.,
\citealt{lo1993,young1996,young1997a,young1997b,young2003});

\item Column 7: The observed rotational velocity of
  the HI component and its uncertainty.  Where available, I quote the
  peak (observed) rotational velocity, else the maximum rotational
  velocity\footnote{For observations that do not extend to
    sufficiently large galactocentric radius, the maximum observed rotational
    velocity will not necessarily equal the peak rotational velocity
    of the galaxy}. In general, I try to give the rotational velocity
  uncorrected for inclination or asymmetric drift, although it is
  sometimes unclear in the original papers if these corrections have
  been applied. ``N/A'' indicates that a rotational signature has been
  looked for and none has been observed. In most cases, it is probably
  safest to interpret the absence of a rotational signature as the
  absence of any rotation with a magnitude equal to or greater than
  the magnitude of the velocity dispersion;

\item Column 8: Dynamical mass within the observed half-light
  radius. I have derived this quantity following the relation of
  \cite{walker2009c}, where $M_{dyn} (\le r_h) = 580\,r_h\,\sigma_\star^2$
  (see also \citealt{wolf2010}) and so values are only given when both
  a {\it stellar} velocity dispersion and a half-light radius are
  available. Here, I adopt the value of the half-light radius measured
  on the semi-major axis. Dynamical masses can, of course, be
  estimated for all systems with measured structures and kinematics,
  but it is beyond the scope of this article to do so. While limited,
  the dynamical mass estimates in Column~8 are at least relatively
  homogeneous, hence comparable;

\item Column 9: References;

\item Column 10: Comments.

\end{itemize}

Figure~8 shows the mass of HI relative to visual luminosity for
galaxies in the Local Group and its immediate vicinity, expressed in
solar units, as a function of distance to either the MW or M31
(whichever is closer). Blue diamonds indicate galaxies with confirmed
HI content. Orange arrows indicate the separations of gas-deficient
galaxies, where the symbols for Sculptor and Fornax indicate that the
presence of HI in these galaxies is ambiguous. Symbol size is
proportional to absolute visual magnitude. The vertical dashed line indicates
the approximate virial radius of the two host galaxies. Also indicated on the top
axis is the ``free-fall time'', that is the time required for the
galaxy to fall from rest at its current position and reach the host galaxy,
acted upon only by the gravity of the giant galaxy,

\begin{equation}
t_{ff} \simeq 14.4\,{\rm Gyr}\,\left(\frac{r}{1\,{\rm Mpc}}\right)^{\frac{3}{2}}\,\left(\frac{M}{10^{12}\,{\rm M_\odot}}\right)^{-\frac{1}{2}}~.
\end{equation}

\noindent Dwarf galaxies with short free-fall time-scales have likely
had many chances to interact with the MW or M31, whereas galaxies with
longer free-fall time-scales likely have not had the time to complete
many orbits. For systems where the free-fall time is of order or
longer than a Hubble time (indicated with a dot-dashed line in
Figure~8), it is reasonable to assume that interactions with giant
galaxies have had a negligible effect on their evolution.

Figure~8 shows the well-noted Local Group position-morphology
relation, first highlighted by \cite{einasto1974}, in which the
gas-deficient dSph galaxies are preferentially found in close
proximity to either the MW or M31. Gas-rich dwarfs are preferentially
located far from either of these two galaxies. Similar behavior has
been noted in other galaxy groupings (e.g.,
\citealt{skillman2003a,geha2006b,bouchard2009} and references therein),
and this has long been interpreted as evidence of the importance of
external effects, such as tidal or ram-pressure stripping, on the
evolution of dwarf galaxies. Indeed, many simulations show that the
combined effects of these processes are in principle sufficient to
remove the gas from dIrr-type galaxies
(e.g., \citealt{mayer2006}). However, \cite{grebel2003} argue that dSph
galaxies are too metal rich for their luminosity in comparison to the
(old) stellar populations in dIrr galaxies; if so, then this requires
that the early metallicity evolution of the two morphological types
(and not just their environments) were also different.

The visual appearance of dIrr galaxies is usually very different from
that of dSph systems, and reflects the presence of very young stellar
populations and on-going star formation in the former. Some dwarf
galaxies are additionally classified as ``dIrr/dSph'': these so-called
transition systems, such as DDO210, distinguish themselves from dIrrs
primarily through the absence of {\it current} star formation
(formally, detectable HI but no detectable HII regions), despite
having very young stellar populations. While star formation in some of
these galaxies may have ceased permanently, in others it may be
possible that low-level star formation is continuing without the
production of HII regions (\citealt{fumigalli2011}). Alternatively,
ongoing, low-level star formation may be interrupted by periods of inactivity
(``gasping''; see \cite{tosi1991} for an example), in which case the
distinction between a dIrr and a transition system merely reflects the
moment we happen to be observing it.

The HST/ANGST survey (\citealt{dalcanton2009}) has recently derived
star formation histories (SFHs) for many nearby dwarf galaxies,
including a large number of the more distant galaxies in this compilation. They show a
tremendous diversity in the SFHs, and highlight that in a statistical
sense there is essentially no difference between the SFHs of dIrr
galaxies and the transition systems (\citealt{weisz2011}). They
suggest that many of the transition dwarfs are perhaps best treated as
low mass dIrr galaxies, and if any of the transition dwarfs are
deserving of the moniker it is likely the less isolated, low gas
fraction systems. \cite{weisz2011} also demonstrate that the
``average'' SFHs of dIrrs and dSphs in their sample are actually
similar over most of cosmic time. It is only within the past few Gyrs,
and particularly within the past 1\,Gyr, that the SFHs
differ. However, the HST/ANGST survey generally samples only bright
dSphs (comparable in luminosity to Fornax, for example), and low
luminosity systems comparable to Ursa Minor, Draco and their ilk are
either absent or under-represented (due to the difficulty of finding
such galaxies outside of the Local Group). The reader is referred to
the HST/LCID survey (\citealt{monelli2010a,monelli2010b,hidalgo2011}
and references therein), that is presently investigating the possible
connections between dIrr, transition and dSph using isolated dwarfs
within the Local Group.

A final distinction between dIrr and dSph galaxies is traditionally
made based on their kinematics, although this is not necessarily well
motivated from an observational perspective, particularly at faint
magnitudes. It is generally argued that dIrr galaxies are rotationally
supported systems, whereas dSph galaxies are pressure supported
systems. However, as visible in Table~4, the majority of dynamical
studies of dSph galaxies are based on resolved spectroscopy of
individual (giant) stars. Until recently, essentially all dynamical
studies of Local Group dIrr galaxies were based on HI, which generally
shows significant rotation. However, recent work by \cite{leaman2009}
has shown that the evolved RGB population in WLM has
$(v_r/\sigma)_\star \simeq 1.2$, whereas $(v_r/\sigma)_{HI} \simeq
6.7$. Thus while it may be true that the stellar component of dIrrs
show significant rotation compared to dSphs, caution should be
exercised in comparing HI kinematics to those of evolved stellar
populations.

Figure~9 plots the stellar velocity dispersion given in Table~4
against the half-light radius for all galaxies for which this
information exists. A broad correlation exists, such that colder
systems have smaller scale-sizes, although the scatter in this
relationship is roughly an order of magnitude at a given
scale-radius. Measuring some of these values is challenging:
\cite{martin2008a} and \cite{munoz2012} both discuss the problems of
determining structural parameters for galaxies in which few
bright stars can be identified (generally the case for galaxies much
fainter than $M_V \simeq -8$). For velocity dispersions, too, the lack
of a large number of bright spectroscopic targets and the low expected
dispersions make these measurements particularly
challenging. Typically, the velocity errors on individual stellar
measurements that are used as a basis for deriving the overall
dispersion are a few km\,s$^{-1}$, which is of order the measured
dispersion for some of the galaxies in Figure~9. One should also
consider that the expected velocity dispersion of a faint system like
Segue, due only to its baryonic mass, is exceptionally small, at
around $\sigma_\star \simeq 0.1$\,km\,s$^{-1}$ (\citealt{mcconnachie2010a}).

There are a couple of primary sources of contamination when measuring
velocity dispersions, all of which have been shown to bias the
measured ``intrinsic'' dispersion of faint galaxies to higher
values. Foreground contamination can have a significant effect. The
first measurement of the velocity dispersion of the Hercules dwarf
gave $\sigma_\star = 5.1 \pm 0.9$\,km\,s$^{-1}$
(\citealt{simon2007}). However, \cite{aden2009a} obtained Stromgren
photometry for stars in the direction of Hercules and were able to
separate member stars from interloping dwarf stars in the halo of the
Galaxy. Without doing this separation, \cite{aden2009a} measure a
value of $\sigma_\star = 7.3 \pm 1.1$\,km\,s$^{-1}$ (in broad with \citealt{simon2007}). By additionally being able to
reject foreground dwarfs, they find a considerably lower value of
$\sigma_\star = 3.7 \pm 0.9$\,km\,s$^{-1}$ (see also
\citealt{aden2009b}). 

Binary stars are a further source of known contamination when
measuring the velocity dispersions of dwarf galaxies, where the reflex
motion of an unresolved binary star is mistaken as a component of the
intrinsic dispersion of the galaxy. This potential source of
contamination was originally investigated prior to the identification
of extremely faint dwarf galaxies to determine if it could inflate the
measured velocity dispersion sufficiently to incorrectly infer the
need for non-baryonic mass-to-light ratios. Numerous studies (e.g.,
\citealt{aaronson1987,mateo1993,olszewski1996,hargreaves1996}) all
concluded that, barring a pathological or extremely high binary population,
binary stars were insufficient to boost a stellar system with
$\sigma_\star \lesssim1$\,km\,s$^{-1}$ to $\sigma_\star \simeq 8 -
10$\,km\,s$^{-1}$, although some inflation of the dispersion is to be
expected. \cite{minor2010} have developed a methodology to
statistically correct the velocity dispersions of dwarf galaxies for this
binary inflation; their own study indicates that systems with intrinsic dispersions
of $4 - 10$\,km\,s$^{-1}$ are unlikely to have their dispersion inflated
by more than 20\,\% as a result of binaries. For systems with lower
intrinsic dispersions than this, \cite{mcconnachie2010a}
argue that the inflation effect due to binaries could be significant,
although this depends on the as-yet-unmeasured stellar binary fraction(s)
in dwarf galaxies. 

Recently, \cite{koposov2011} develop a procedure for obtaining
accurate radial velocities with demonstrably reliable uncertainties at
faint magnitudes and conduct a multi-epoch spectroscopic study of the
Bootes dwarf galaxy. Approximately 10\% of the stars in their sample
show evidence for velocity variability. Their data favor a two
component model for Bootes (see discussion in \S4), where the central, dominant
component has a velocity dispersion of $\sigma_\star =
2.4^{+0.9}_{-0.5}$\,km\,s$^{-1}$. Their preferred value is notably
lower than earlier values of order $\sigma_\star \simeq
6$\,km\,s$^{-1}$ reported by \cite{munoz2006b} and \cite{martin2007b}
(although they are in statistical agreement).  These results all
emphasize the difficulties in obtaining reliable estimates of the
intrinsic velocity dispersions of very faint dwarf galaxies: reliable
(and small!) velocity uncertainties, multi-epoch data and dwarf-giant
separation are all essential components to any successful measurement.

With the above caveats on the basic dynamical data, Figure~10 shows
the dynamical mass estimates for each dwarf galaxy within its
half-light radius, as a function of absolute visual magnitude. Here, I
have used the formalism by \cite{walker2009c}; \cite{wolf2010} justify
in detail why the mass within the half-light radius is in general able
to be reliably measured for dispersion-supported spheroids. The left
panel highlights by name each galaxy for which the necessary data
exist, whereas the right panel shows the formal uncertainties on
propagating the quoted uncertainties on $\sigma_\star$ and $r_h$. Note
that these estimates do not take into consideration errors introduced
as a result of the dynamical state of the galaxy. For example, a few
galaxies in Figure~10 show some rotation in addition to dispersion
support (M32, NGC205, NGC147, NGC185, Cetus, WLM) and other galaxies
may or may not be in dynamical equilibrium. Further, the Walker et
al. formalism assumes that the velocity dispersion profile is flat
with radius (the Wolf et al. formalism adopts a luminosity-weighted
mean velocity dispersion); however, for many galaxies shown in
Figure~10 the form of the radial velocity dispersion profile is
unknown. A full study of the dynamics of these galaxies is clearly
outside the scope of this article, and the current estimates are
provided solely to demonstrate the status of the observational data in
terms of its basic implications regarding the mass content of the
galaxies. Figure~11 shows the same data as Figure~10, except I have
now normalised $M_{dyn}$ by $0.5\,L_\star$ to estimate the implied
mass-to-light ratio of each system within its half-light radius.

Figures~10 and 11 demonstrate that there appears to be a well 
defined relation consistent with a power law between $M_{dyn} (\le r_h)$ and $M_V$ that
holds across a factor of nearly 1 million in luminosity, in agreement
with the finding reported by \cite{walker2009c}. As discussed above,
systematic errors in the measurements of the relevant quantities, and
the necessary assumptions that must be made in converting these
quantities to a mass estimate, mean the precise position of any single
point in these plots is open to considerable debate. Further, it
should be stressed that observational selection effects such as that
discussed for Figure~6 (which may be responsible for the precise form
of the $M_V - r_h$ at low luminosities) will also affect these
figures due to the dependence of $M_{dyn}$ on $r_h$.

Measurements such as those on which Figures~10 and 11 are based have
also been used to argue for a common mass-scale for the (Milky Way)
dwarfs (\citealt{strigari2007b}). However, this is usually made with
reference to the mass enclosed with an absolute radial scale
(typically $r= 300$\,pc), a quantity that is used because of its
usefulness in dark-matter only simulations, where there is no
information regarding the size of the baryonic components of galaxies.
However, given that the half-light radii of the dwarfs in and around
the Local Group have $ 20 \lesssim r_h \lesssim 2000$\,pc, then from
an observational perspective this is a far less well-defined quantity
than $M_{dyn} (\le r_h)$.

\section{Mean stellar metallicities}

Table~5 lists the available data for the mean stellar metallicities, $<[Fe/H]>$, of the galaxies:

\begin{itemize}

\item Column 1: Galaxy name;

\item Column 2: Mean stellar metallicity and error, generally derived
  from observations of mostly evolved giants, else commented on in
  Column 5. Note that I quote the {\it mean} metallicity of all stars
  (as opposed to the median, mode, etc.) in the observed metallicity
  distribution function, and so this will be susceptible to any
  uncorrected biases in the observed sample. The mean stellar metallicity refers to the
  mean iron-peak metallicity expressed in logarithmic notation
  relative to the solar value;

\item Column 3: Technique used to estimate stellar metallicity. A
  discussion is given below;

\item Column 4: References;

\item Column 5: Comments.

\end{itemize}

Figure~12 shows mean stellar metallicity as a function of absolute visual
magnitude, where by definition both axes are logarithmic
quantities. In the left panel, all tabulated data from Table~5 are
given, whereas in the right panel only data that use
spectroscopically-determined quantities are shown. 

Generally the most well-accepted method of determining [Fe/H] for a
given star is from direct high resolution spectroscopy (HRS) from which the
strengths of iron lines can be measured directly. However, Table~5
lists estimates for $<[Fe/H]>$ that stem from a variety of additional techniques:

\begin{itemize}

\item RGB color: \cite{dacosta1990} derive a relation between the mean
  color of RGB stars in globular clusters measured at $M_V = -3$ and
  the mean stellar metallicity of the cluster. A refined calibration
  (derived at $M_V = -3.5$) using the same data was presented by
  \cite{lee1993a}, and this has been used throughout the literature as
  a convenient means of estimating the mean stellar metallicity of a system
  when only photometry of bright stars is available. An obvious
  disadvantage, of course, is that it is calibrated on ancient stellar
  populations. Since RGB stars are subject to age and metallicity
  degeneracies then this technique may not be reliable if
  significantly younger stars are present. I note that in Table~5 and
  Figure~12 most of the non-Local Group galaxies have their mean stellar
  metallicity determined in this way. There is a suggestion in
  Figure~12 that these estimates are low in comparison to Local Group
  galaxies with similar absolute visual magnitudes, but its cause or
  significance is difficult to determine;

\item Isochrones: The comparison of theoretical isochrones,
  evolutionary tracks, or globular cluster fiducials, to the CMDs of
  galaxies is usually heavily or entirely weighted towards the color
  of the RGB, and so is in principle subject to the same systematic
  uncertainties regarding ages as discussed
  above. In addition, isochrones produced by different groups may have
  some systematic differences between the predicted RGB colors at any
  age. However, one potential advantage of isochrones over empirical
  RGB color relations is that, if additional information is available
  regarding ages (e.g., from the red clump or main sequence turn-off),
  then isochrones with appropriate ages can be used;

\item CMD fitting: The reader is referred to the many excellent papers
  on SFH analysis using resolved stellar populations for more details
  (e.g., \citealt{gallart2005,tolstoy2009} and references
  therein). Here, age degeneracies will likely be much reduced since
  age information is available from elsewhere in the CMD (in
  particular the main-sequence turn-off region);

\item Calcium II Triplet equivalent widths (low, moderate resolution
  spectroscopy): The strengths of the three CaT absorption lines at
  $\lambda = 8498, 8542$ and $8662$\,\AA in RGB stars have been shown to
  correlate with [Fe/H] if the temperature and surface gravity of the
  star is able to be estimated through measurement of its absolute
  magnitude. The CaT is a relatively strong feature and consequently it
  can be relatively easily measured using low-to-moderate resolution
  spectroscopy (L/MRS). Many authors (e.g., \citealt{rutledge1997}) have
  developed calibrations between CaT equivalent width and
  [Fe/H], most recently \cite{starkenburg2010}. These authors also
  examine the form of the correlation at [Fe/H] $\lesssim -2.5$. Using
  the new calibration, \cite{starkenburg2010} demonstrate that the resulting
  estimates of [Fe/H] match those derived from HRS down to a limiting
  stellar metallicity of [Fe/H] $\simeq -4$. Many of the values listed in
  Table~5 use the earlier calibrations, which can potentially deviate from the new
  calibration for stellar metallicities lower than [Fe/H] $\simeq -2$,
  depending upon the absolute magnitudes of the observed stars;

\item Spectral synthesis (moderate resolution spectroscopy):
  Here, synthetic spectra spanning a range of key parameters (e.g.,
  [Fe/H], [$\alpha/$Fe], $T_{eff}$, $\log{g}$) are compared to the
  measured stellar spectrum to determine the most likely
  [Fe/H]. \cite{kirby2008a} have developed this technique for specific
  use on the dwarf galaxies of the MW with considerable success. Here,
  the main advantages are that one uses more spectral information than
  relying only on specific indices. Since one uses moderate
  resolutions, fainter targets can be studied than is possible with HRS.

\end{itemize}

Clearly, the range of techniques from which $<[Fe/H]>$ has been
derived in Table~5 may produce numerous systematic differences between
results. Recent discussions of some of these issues can be found in
\cite{kirby2008a} (spectral synthesis), \cite{starkenburg2010} (CaT)
and \cite{lianou2011} (isochrones, CaT) and the reader is referred to
these papers for more details. I note that several of the $<[Fe/H]>$
values listed in Table~5 only quote the uncertainty from the random
error in the mean (i.e., $\frac{\sigma_{[Fe/H]}}{\sqrt{N}}$) and do
not include the additional systematic uncertainties (which cannot be
treated in the same way). For galaxies for which this is the case, an
additional systematic error component must be included, typically a
tenth of a dex or more.

In addition to the systematic uncertainties introduced by the
measurement technique, an additional systematic effect enters the
calculation of the mean stellar metallicities in Table~5 through the
spatial extent of the stars contributing to the measurement. Dwarf
galaxies are known to exhibit radial gradients in stellar populations,
ages and/or metallicity distributions (e.g.,
\citealt{harbeck2001,tolstoy2004,battaglia2006,bernard2008} and
references therein). Thus far, all observed metallicity gradients are such that a
higher ``mean'' stellar metallicity is observed in the central regions
of the dwarf than in the outer regions. Two galaxies with identical
metallicity distribution functions and radial gradients will therefore
be measured to have different ``mean'' stellar metallicities if there
is any difference in the radial distribution of stars that contribute
to the measurement. Given the range in radial scale associated with
the galaxies in Table~5, and the unknown spatial structure of the
stellar populations/ages/metallicities within many of these systems,
this effect may be significant.

With the above caveats, both panels of Figure~12 show the well-known
luminosity -- mean stellar metallicity relation, whereby more luminous
galaxies are generally more metal-rich in the mean than faint galaxies
(see also \cite{tremonti2004,lee2006,kirby2008b} and references
therein). The relation is better defined in the right panel, which
uses only spectroscopic measurements of $<[Fe/H]>$; the degree to
which the scatter in the left panel is a result of systematic errors
(age effects, measurement techniques, etc.) is unknown. As pointed out
by \cite{kirby2008b}, the luminosity -- mean stellar metallicity
relation at faint magnitudes ($M_V > -8$) is entirely consistent with
a continuation of the trend displayed for bright galaxies. However, it
is also worth highlighting the dearth of data at faint magnitudes: if
only galaxies fainter than $M_V = -8$ are considered, then the data
are equally consistent with a scatter around a floor in mean stellar
metallicity at $<[Fe/H]>$ $\simeq -2.1$. While more data are certainly
needed, it is interesting to note that the magnitude at which one can
start arguing for a floor in mean stellar metallicity is broadly the
same point at which there is a break in the absolute magnitude --
surface-brightness relation shown in Figure~7  (discussed in
\S4). While certainly speculative, this may suggest a direct link
between the surface brightness of a galaxy (specifically, the {\it
  density} of baryons, rather than the total amount of baryons) and
its mean stellar metallicity (see theoretical work by \cite{dekel2003},
\cite{revaz2012} and references therein; see also
\citealt{skillman2003b}).

Many more parameters are required in order to fully define the
metallicity distribution function of stars in any of the dwarf
galaxies, and the reader is referred to \cite{tolstoy2009} for a more
complete discussion of ongoing observational efforts in this
regard. As discussed in \cite{tolstoy2009}, detailed chemistry, such
as abundance ratios of key elements, are now available for the nearest
galaxies, with much more to come. In the near future, the combination
of dynamical data with detailed information on the chemical evolution
of galaxies is likely to continue to be a significant area of
discovery and advance.

\section{Summary}

Tables~1 -- 5 provide observational parameters for approximately 100
nearby dwarf galaxies, including positional data, radial velocities,
photometric and structural parameters, stellar, HI and dynamical
masses, internal kinematics (dispersions and rotation), and mean
stellar metallicities. In addition to discussing the provenance of the
values and possible sources of uncertainties affecting their usage, I
also consider the membership and (limited?) spatial extent of the MW
and M31 sub-groups; the morphological diversity of the M31 sub-group
in comparison to the MW; the location of the zero-velocity surface of
the Local Group; the time-scales that can be associated with the
orbital/interaction histories of the Local Group members; the scaling
relations defined by the sample, including their behavior at the
faintest magnitudes/surface brightnesses; the luminosity -- mean
stellar metallicity relation, including its possible connection to the central
surface brightness of the galaxies.

The dwarf galaxies discussed herein constitute all known satellite
systems in the surrounding of the MW and M31, all quasi-isolated
systems in the more distant reaches of the Local Group, and all other,
generally very isolated, dwarf galaxies that surround the Local Group
out to 3\,Mpc, beyond which most of the known galaxies are members of
Maffei, Sculptor, Canes Venatici, IC342, M81 and the many other groups
in the Local Volume found out to the Virgo Cluster and beyond. In some
respects, these systems are stepping stones into the Local Volume, and
we are fortunate that they present us with a large and diverse range
of properties to explore. This article has purposefully focused
entirely on observable, quantifiable properties of these galaxies, and
it is right that the published tables likely already contain obsolete
numbers or do not include the most recent discoveries\footnote{Updated
  versions of the core data available in Tables 1 -- 5 for all dwarf
  galaxies within 3\,Mpc are available at
  https://www.astrosci.ca/users/alan/Nearby\_Dwarfs\_Database.html}. Our knowledge of many
of these systems far exceeds that of other galaxies, and we will
continue to be able to obtain ever more detailed observations of
aspects of their structure, dynamics, stellar populations and chemical
signatures that are beyond the reach of observational programs of more
distant galaxies. In the future as now, these are the galaxies for
which we will know the most, and so are perhaps destined to understand
the least.

\acknowledgements Updated versions of the core data available in
Tables 1 -- 5 for all dwarf galaxies within 3\,Mpc are available at
https://www.astrosci.ca/users/alan/Nearby\_Dwarfs\_Database.html. I thank Peter Stetson,
Matthew Walker, Daniel Weisz, Sidney van den Bergh and Else
Starkenburg for careful readings of the manuscript and valuable
comments; Ryan Leaman for his feedback, and for stressing the
importance of radial metallicity gradients when considering mean
stellar metallicities, which led to the addition of the relevant
paragraph; Josh Simon for highlighting relevant references. In
addition, I am grateful to Vasily Belokurov, Michelle Collins, Patrick
C{\^o}t{\'e}, Stephanie C{\^o}t{\'e}, Julianne Dalcanton, Tim Davidge,
Marla Geha, Karoline Gilbert, Carl Grillmair, Mike Irwin, Bradley
Jacobs, Ryan Leaman, Nicolas Martin, Emma Ryan-Weber, Evan Skillman,
Peter Stetson, Brent Tully, Sidney van den Bergh, Matthew Walker,
Daniel Weisz and Lisa Young for their help in answering questions that
came up during the preparation of this manuscript; to Wolfgang
Steinicke and Harold Corwin Jr. for their invaluable assistance in
tracing some of the early references relating to the original
discoveries of these objects; to the referees for their supportive
comments and very helpful feedback that led to significant
improvements in the text. Finally, thanks go to the organizers of the
KITP program ``First Galaxies and Faint Dwarfs: Clues to the Small
Scale Structure of Cold Dark Matter'', at which this article was
completed. As such, this research was supported in part by the National
Science Foundation under Grant No. NSF PHY11-25915


\clearpage

\begin{figure*}
  \begin{center}
    \includegraphics[angle=270, width=13cm]{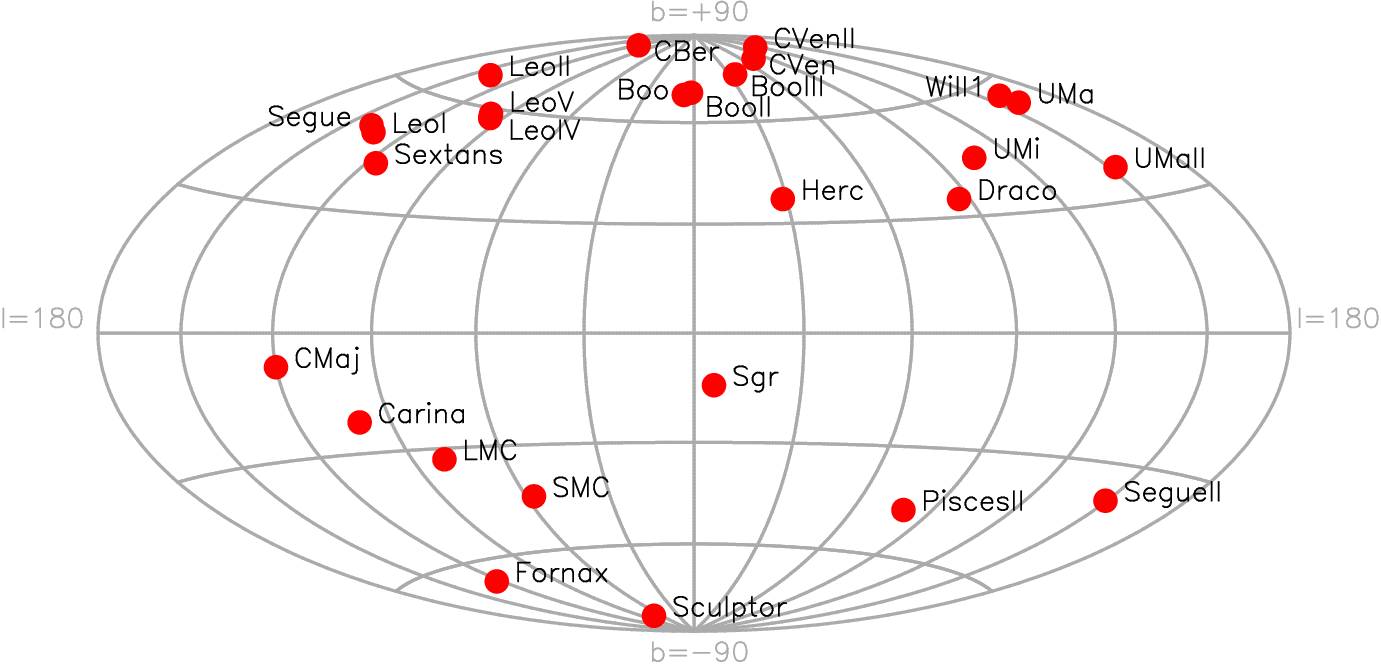}
    \includegraphics[angle=270, width=13cm]{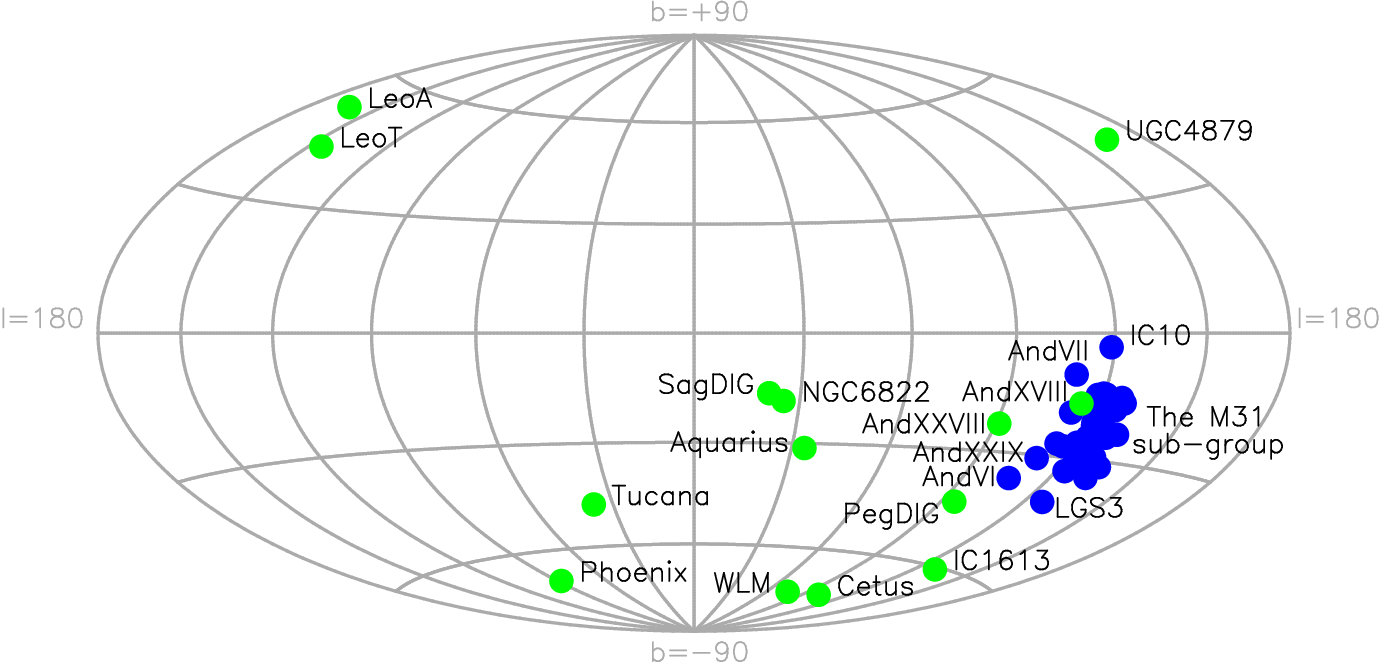}
    \includegraphics[angle=270, width=13cm]{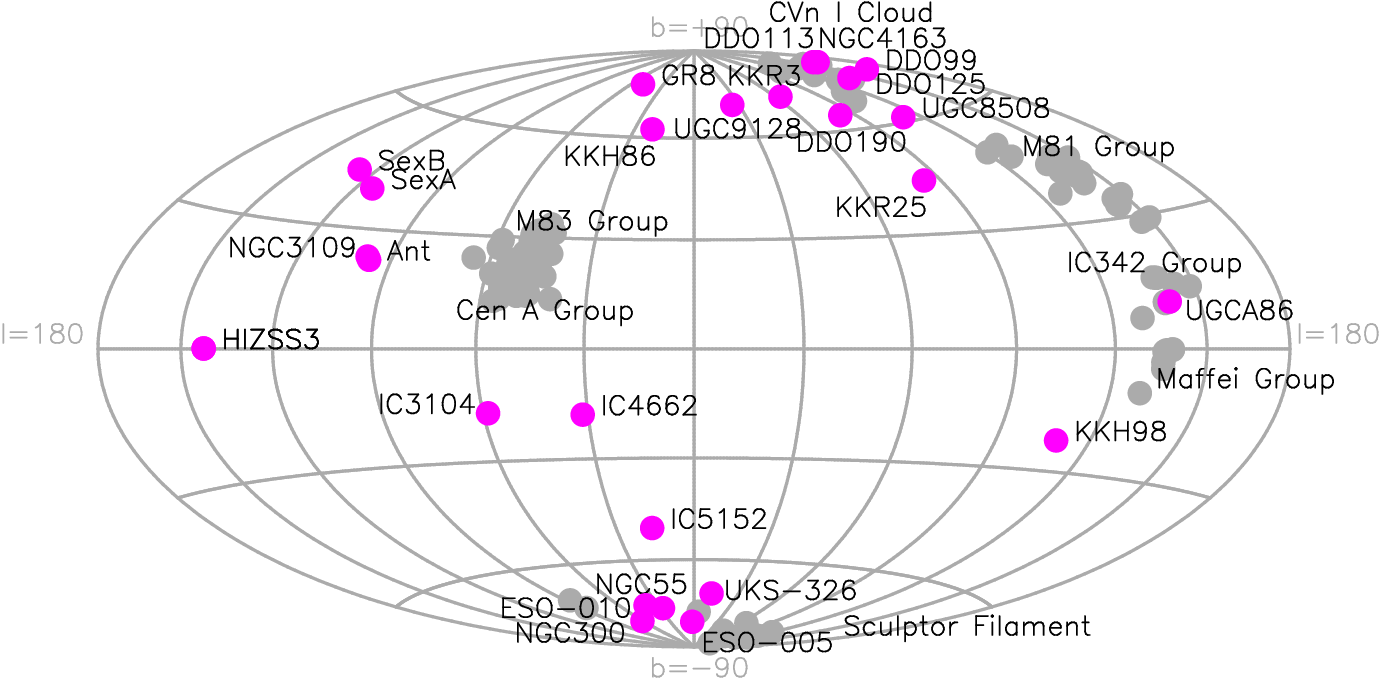}
   \caption{{\footnotesize Aitoff projections of the Galactic coordinates of MW
     galactic satellites (top panel); the M31 sub-group (blue) and
     isolated Local Group galaxies (green; middle panel);
     the nearest galaxies to the Local Group that have distances based
     on resolved stellar populations that place them within 3\,Mpc
     (magenta; bottom panel). The positions of nearby
     galaxy groups are indicated in grey in the bottom panel.}}
  \end{center}
\end{figure*}

\clearpage

\begin{figure*}
  \begin{center}
    \includegraphics[angle=270, width=12cm]{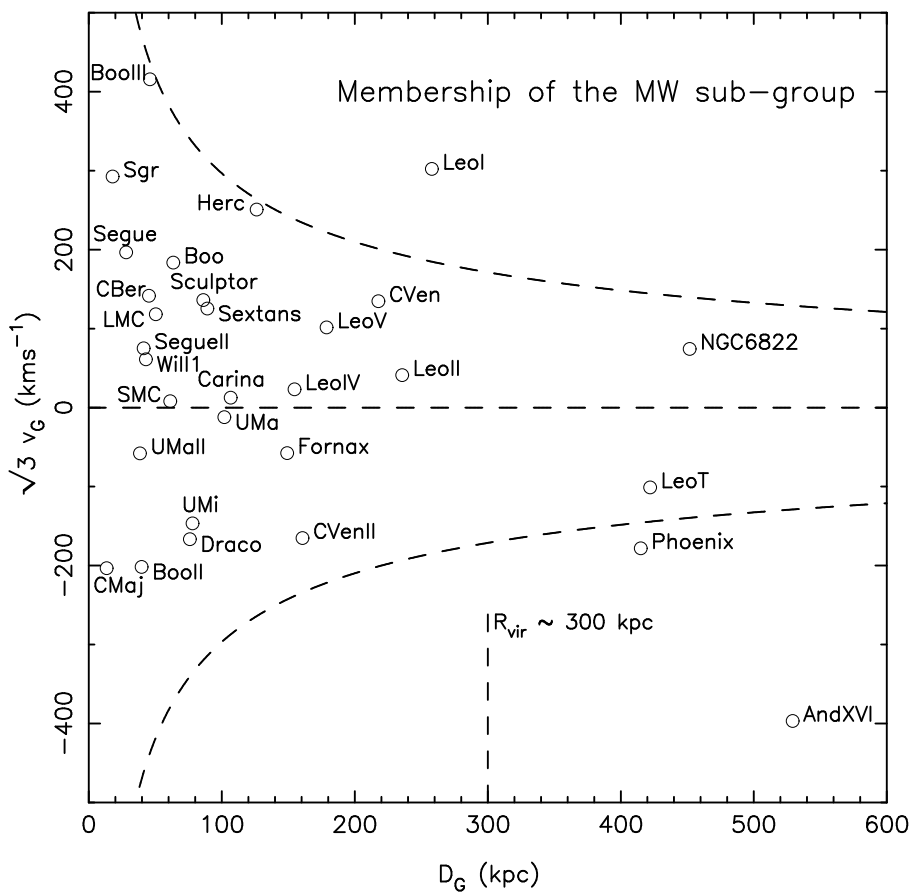}
   \caption{Galactocentric velocity versus distance for all galaxies
     in the proximity of the Milky Way. The y-ordinate has been
     multiplied by a factor of $\sqrt3$ to account for the unknown
     tangential motions of the galaxies. Dashed curves show the escape
   velocity from a $10^{12}\,M_\odot$ point-mass. The vertical dashed
   line indicates the approximate location of the expected
   cosmological virial
   radius of the Milky Way ($R_{vir} \sim 300$\,kpc).}
  \end{center}
\end{figure*}

\clearpage

\begin{figure*}
  \begin{center}
    \includegraphics[angle=270, width=12cm]{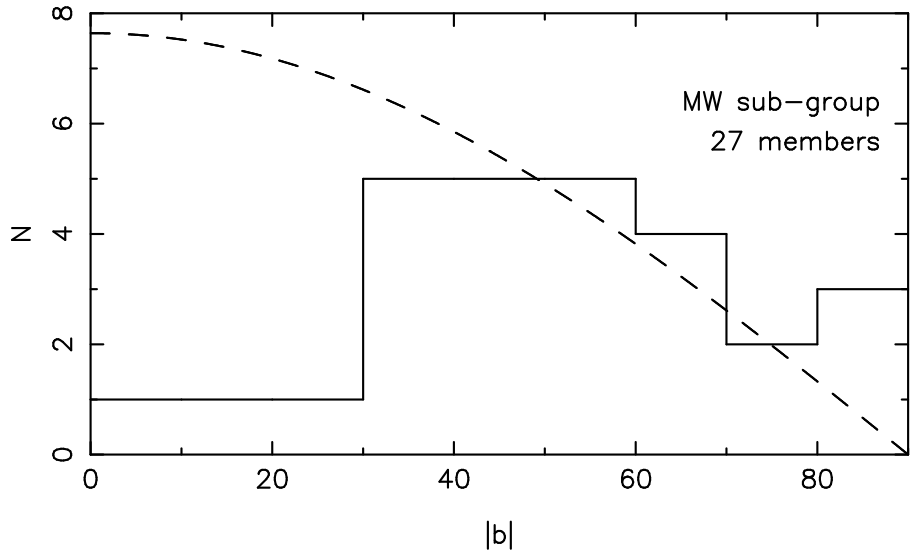}
  \caption{Galactic latitude distribution of Milky Way dwarf
    galaxies. The dashed curve shows an isotropic distribution
    normalised to match the current number of dwarfs known at
    $|b|>30^\circ$ (24), and highlights the dearth of objects at low
    Galactic latitude.}
  \end{center}
\end{figure*}

\clearpage

\begin{figure*}
  \begin{center}
    \includegraphics[angle=270, width=12cm]{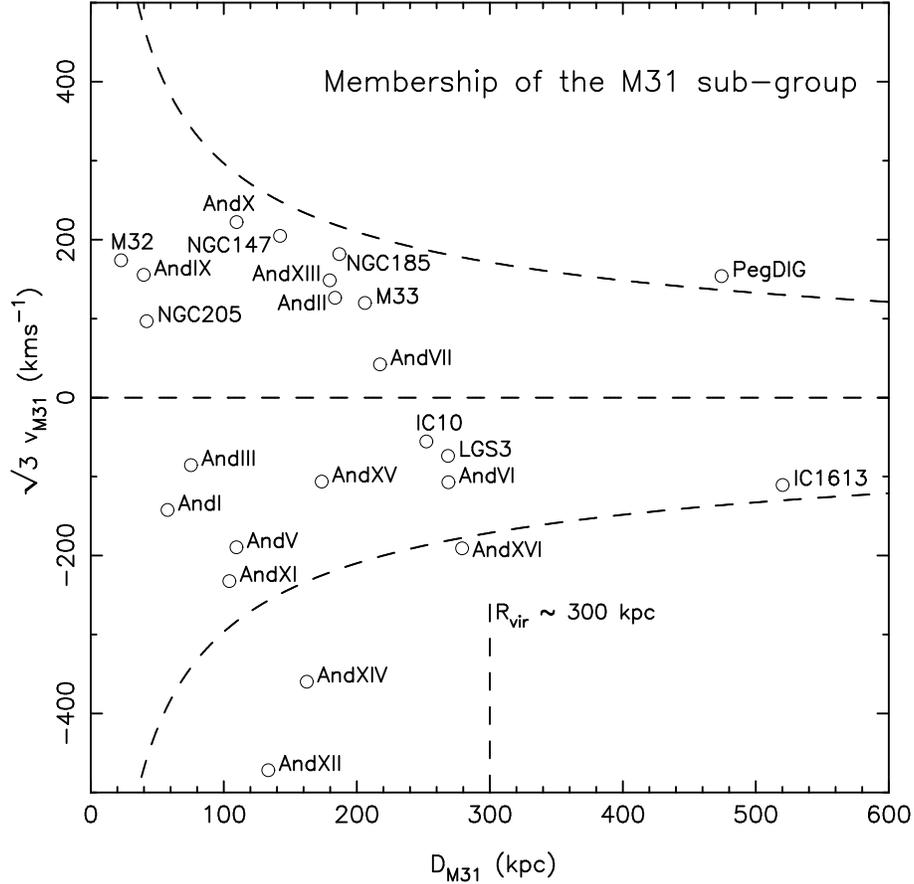}
   \caption{M31-centric velocity versus distance for all galaxies
     in the proximity of M31. The y-ordinate has been
     multiplied by a factor of $\sqrt3$ to account for the unknown
     tangential motions of the galaxies. Dashed curves show the escape
   velocity from a $10^{12}\,M_\odot$ point-mass. The vertical dashed
   line indicates the approximate location of the expected
   cosmological virial
   radius of M31 ($R_{vir} \sim 300$\,kpc).}
  \end{center}
\end{figure*}

\clearpage

\begin{figure*}
  \begin{center}
    \includegraphics[angle=270, width=16cm]{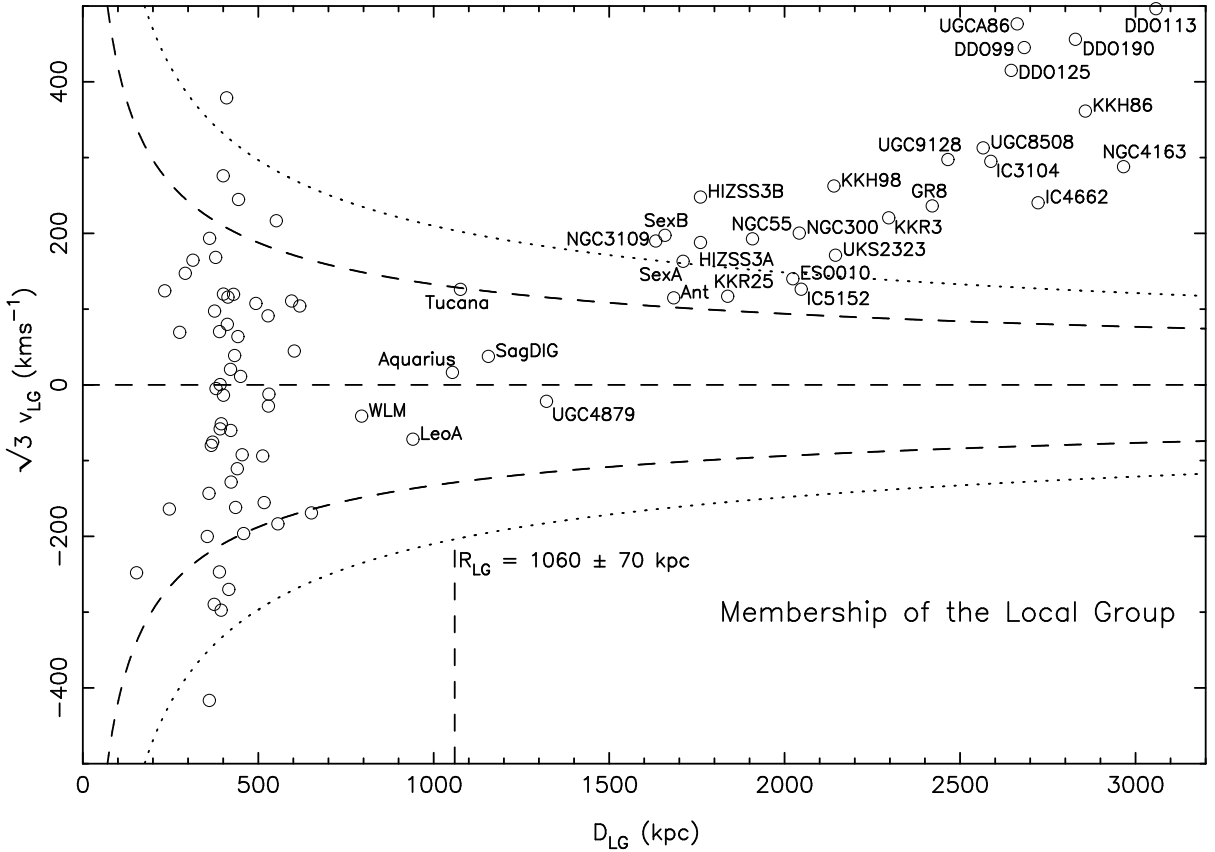}
   \caption{Local Group-centric velocity versus distance for all galaxies
     in the proximity of the Local Group. The y-ordinate has been
     multiplied by a factor of $\sqrt3$ to account for the unknown
     tangential motions of the galaxies. Dashed curves show the escape
   velocity from a $2 \times 10^{12}\,M_\odot$ point-mass. Dotted
   curves show the escape velocity from a $5 \times 10^{12}\,M_\odot$ point-mass,
   as implied from the timing argument (\citealt{lyndenbell1981}). The vertical dashed
   line indicates the zero-velocity surface of the Local
   Group ($R_{LG} = 1060 \pm 70$\,kpc), derived here from the mean
   distance of the six  highlighted galaxies that cluster around zero velocity.}
  \end{center}
\end{figure*}

\clearpage

\begin{figure*}
  \begin{center}
    \includegraphics[angle=270, width=10cm]{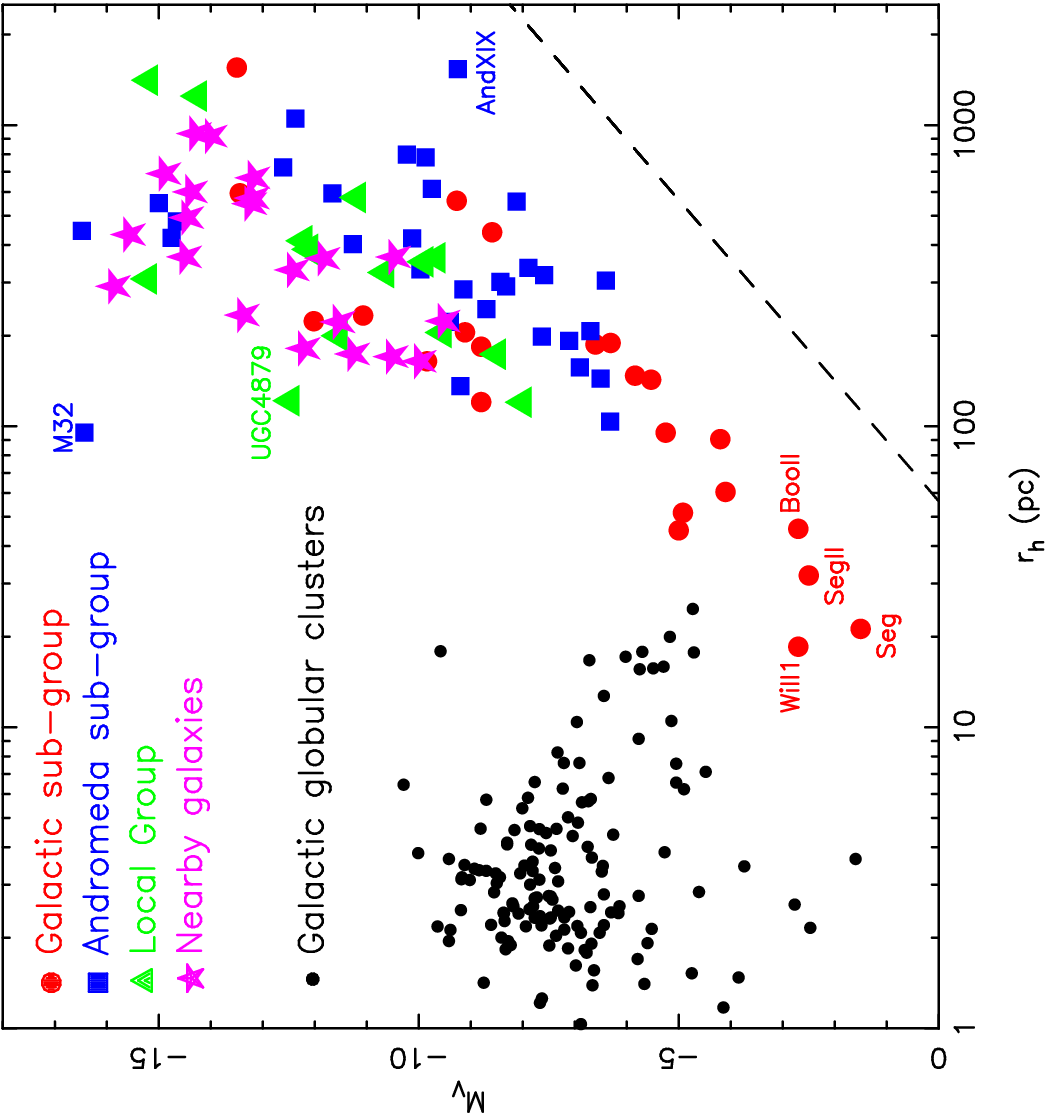}
    \includegraphics[angle=270, width=10cm]{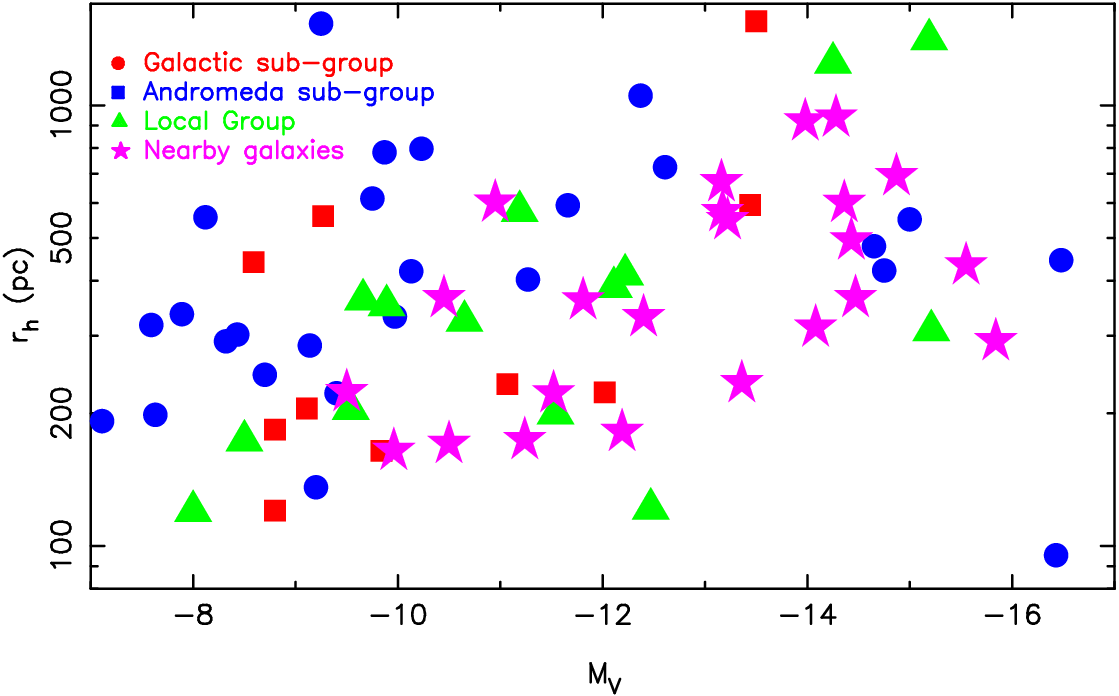}
    \caption{Top panel: The absolute visual magnitude versus
      half-light radius (geometric mean of the major and minor axes)
      for the galaxy sample (where the symbols and color-code identify
      Galactic, M31, Local Group and nearby galaxies, as explained
      in the key). Also indicated as small black dots are the location
      of the Galactic globular clusters, using the data compiled by
      \cite{harris1996}. Obvious outliers to the overall trend
      displayed by the galaxies are highlighted by name, as are some
      of the least luminous (candidate) dwarf galaxies. The dashed
      line indicates the direction defined by points of constant
      surface brightness (averaged within the half-light
      radius). Bottom panel: Same parameters as top panel showing an
      enlargement around $-17\le M_V\le -7$.}
  \end{center}
\end{figure*}

\clearpage

\begin{figure*}
  \begin{center}
    \includegraphics[angle=270, width=12cm]{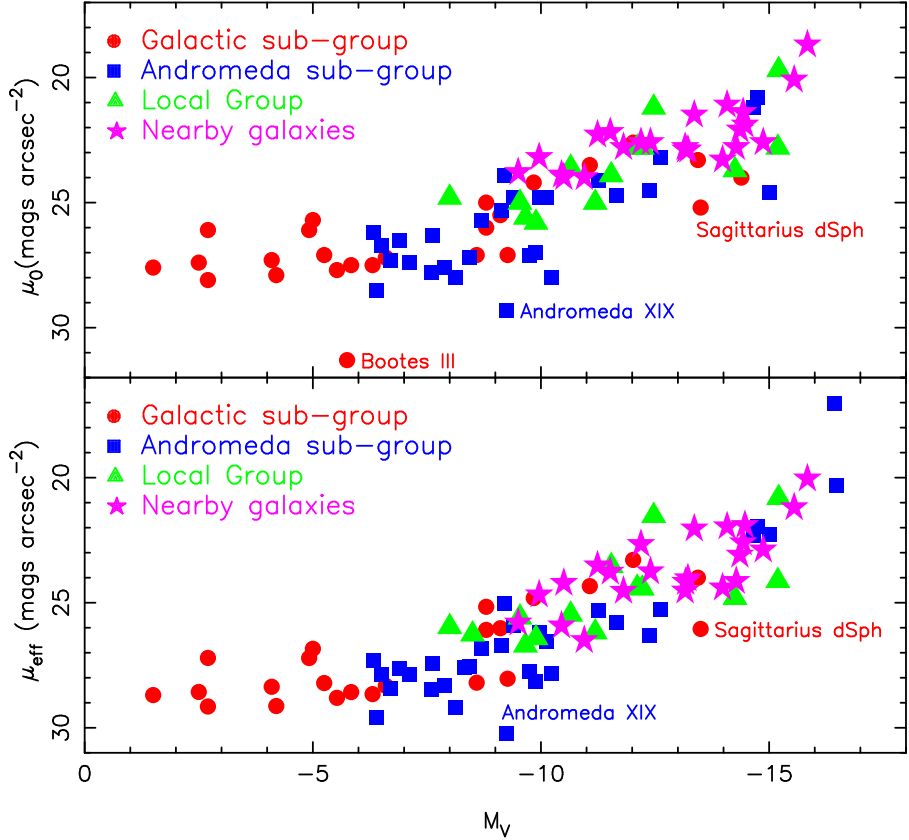}
    \caption{Surface brightness versus absolute visual magnitude for
      the galaxy sample, where symbols have the same meaning as in
      Figure~6. The top panel uses central surface brightness, whereas
      the bottom panel uses the surface brightness averaged over the
      half-light radius. Obvious outliers to the overall trends are
      highlighted by name. Note that Bootes III is not present in the
      lower panel due to the lack of a measured value for $r_h$.}
  \end{center}
\end{figure*}

\clearpage

\begin{figure*}
  \begin{center}
    \includegraphics[angle=270, width=16cm]{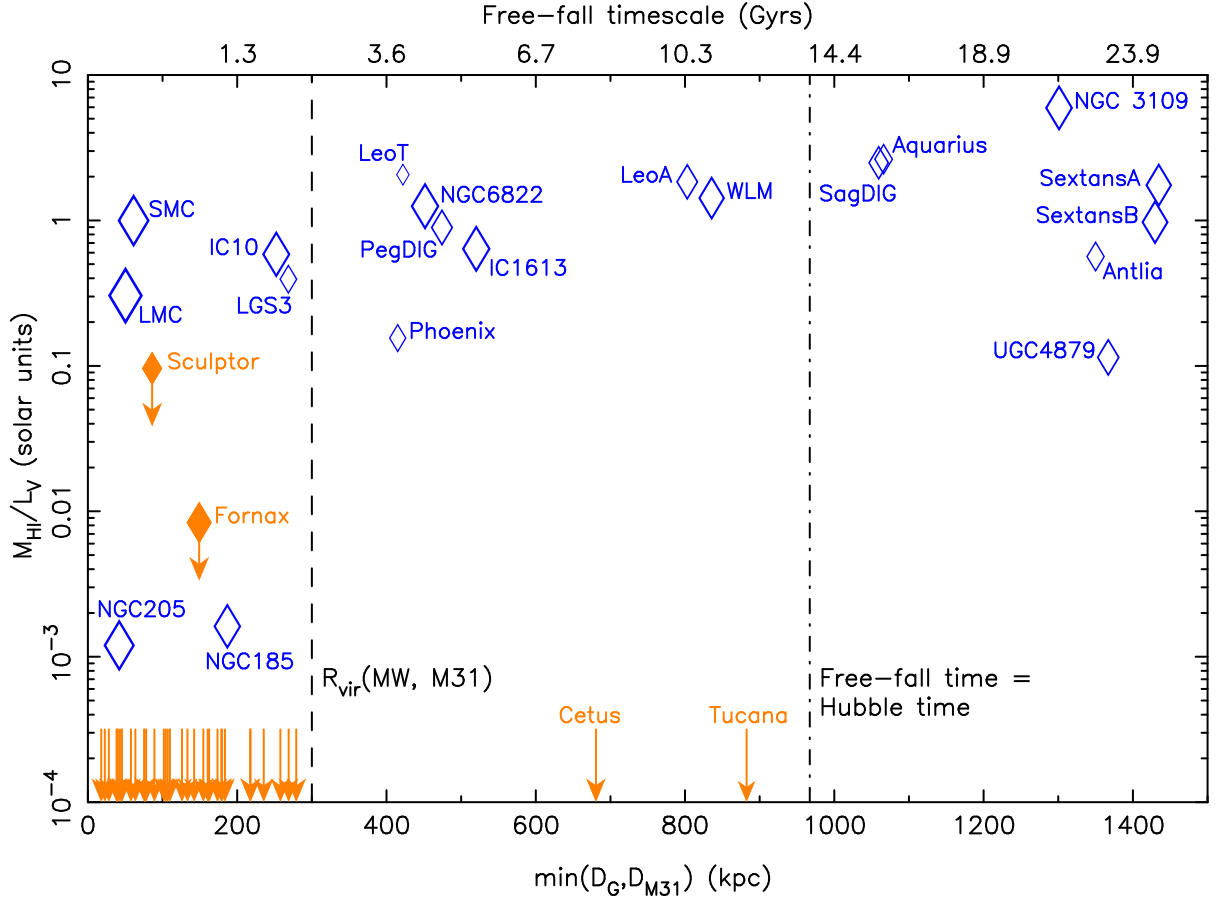}
    \caption{HI fraction, expressed as $M_{HI}/L_V$ in solar units, as
      a function of proximity to a giant galaxy. Blue diamonds
      indicate galaxies with confirmed HI content. Orange arrows
      indicate the separation of gas-deficient galaxies from either
      the MW or M31. The symbols for Sculptor and Fornax indicate that
      the presence of HI in these galaxies is ambiguous. Symbol size
      is proportional to absolute visual magnitude. The vertical dashed line
      indicates the approximate virial radius of the MW/M31. Also
      indicated on the top axis is the time required for the galaxy to
      accelerate from rest and reach the MW/M31. The dot-dashed line
      indicates the separation at which this time equals a Hubble
      time.}
  \end{center}
\end{figure*}

\clearpage

\begin{figure*}
  \begin{center}
    \includegraphics[angle=270, width=12cm]{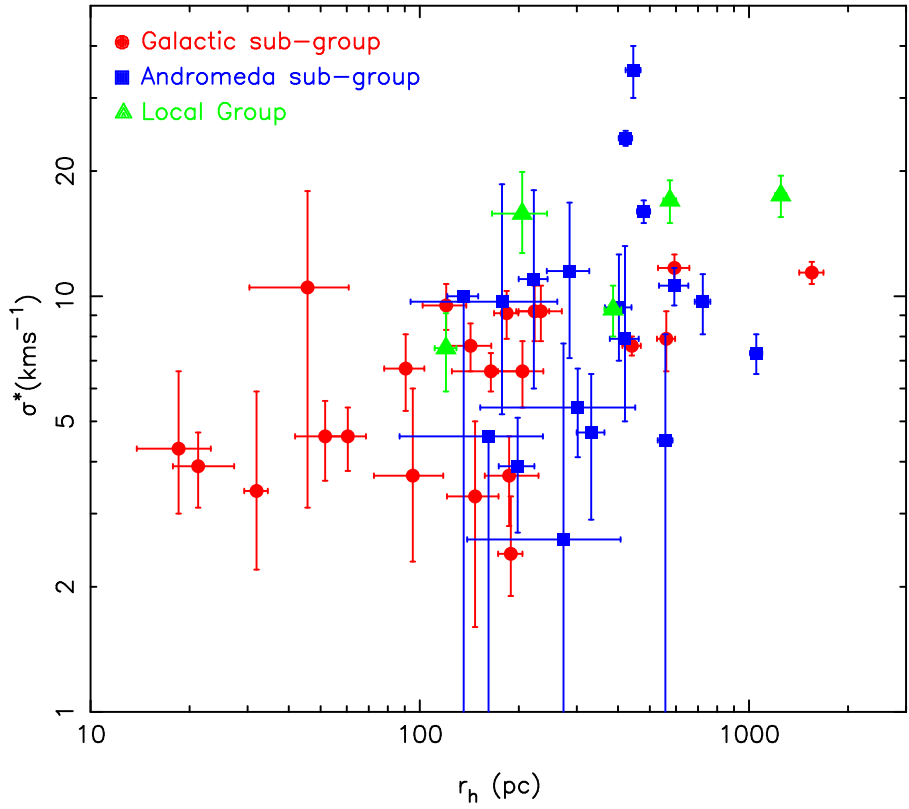}
    \caption{Stellar velocity dispersion versus the (geometric-mean)
      half-light radius measured for all galaxies for which this
      information is available.}
  \end{center}
\end{figure*}

\clearpage

\begin{figure*}
  \begin{center}
    \includegraphics[angle=270, width=8cm]{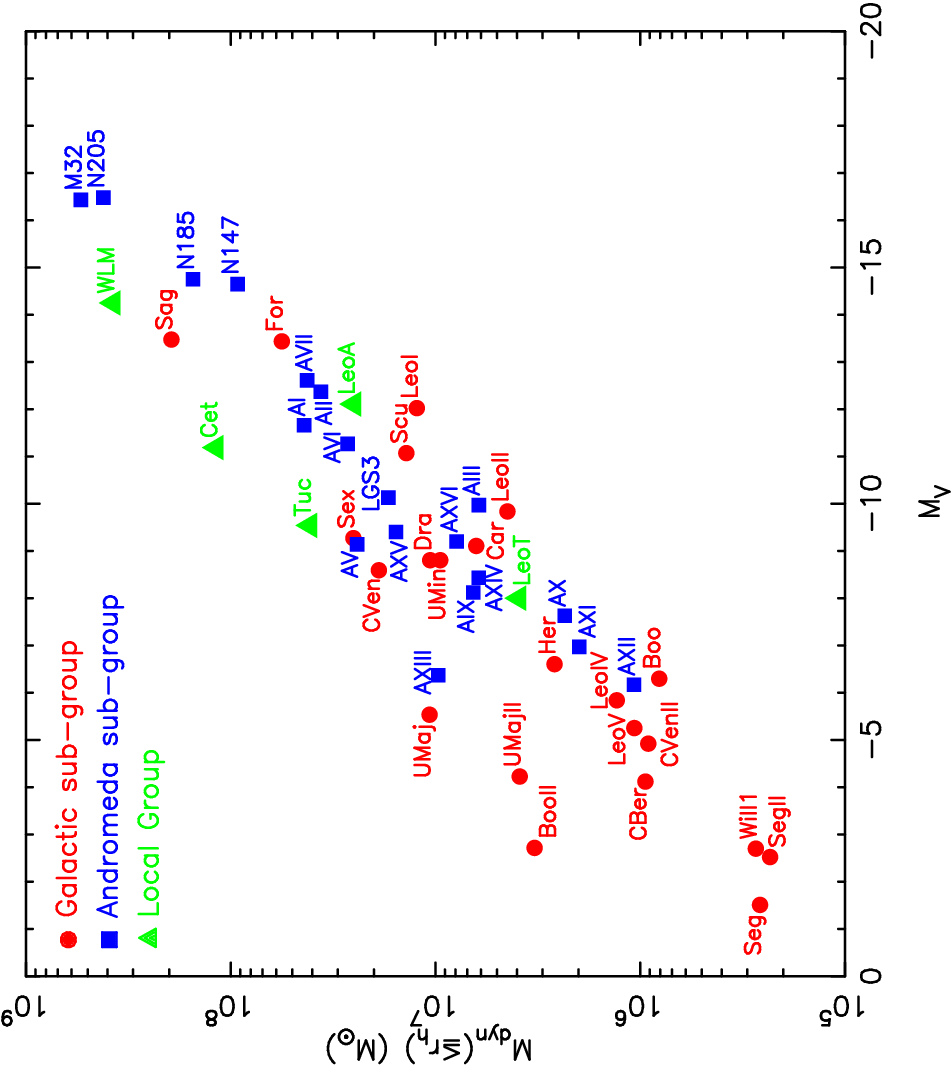}
    \includegraphics[angle=270, width=8cm]{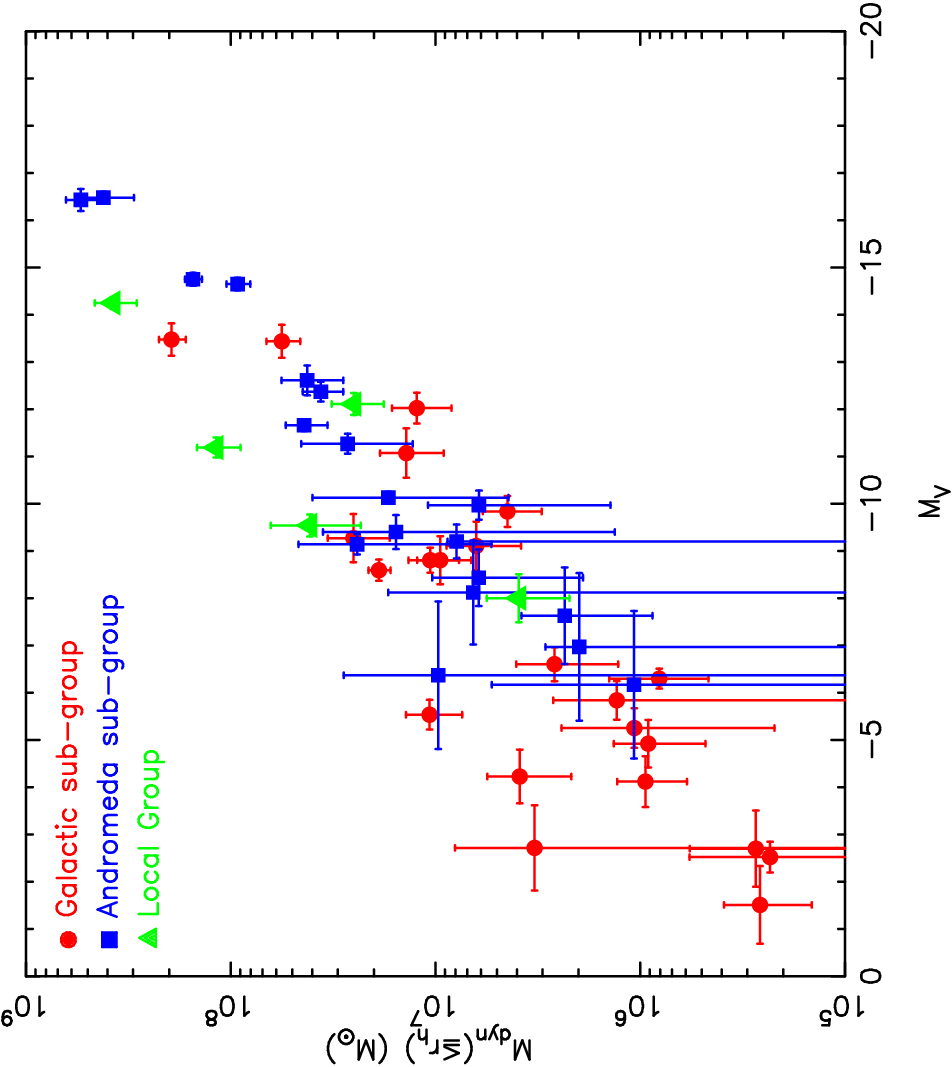}
    \caption{Dynamical mass estimates within the half-light radius
      versus absolute visual magnitude, for
      all galaxies for which stellar velocity dispersion and
      half-light radius information is available. These have been
      calculated using the relation from \cite{walker2009c}. Individual galaxies
      are identified in the left panel, and error bars are
      indicated in the right panel. Note that the errors were
      calculated by propagating the uncertainties on the stellar
      velocity dispersion and half-light radius in the usual way, and
      do not account for any systematic uncertainties in either these
      quantities or in the dynamical state of the galaxy.}
  \end{center}
\end{figure*}

\clearpage

\begin{figure*}
  \begin{center}
    \includegraphics[angle=270, width=12cm]{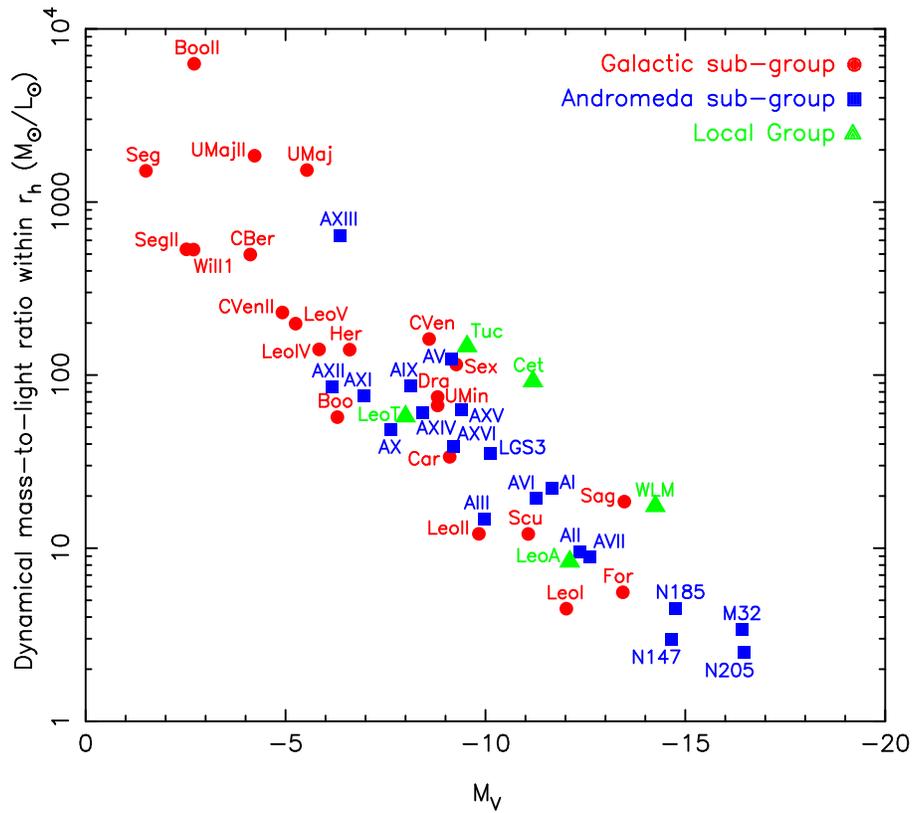}
    \caption{Mass-to-light ratio, in solar units, calculated within the half-light
      radius for all dwarf galaxies for which the necessary data
      exist, as a function of absolute visual magnitude.}
  \end{center}
\end{figure*}

\newpage

\begin{figure*}
  \begin{center}
    \includegraphics[angle=270, width=8cm]{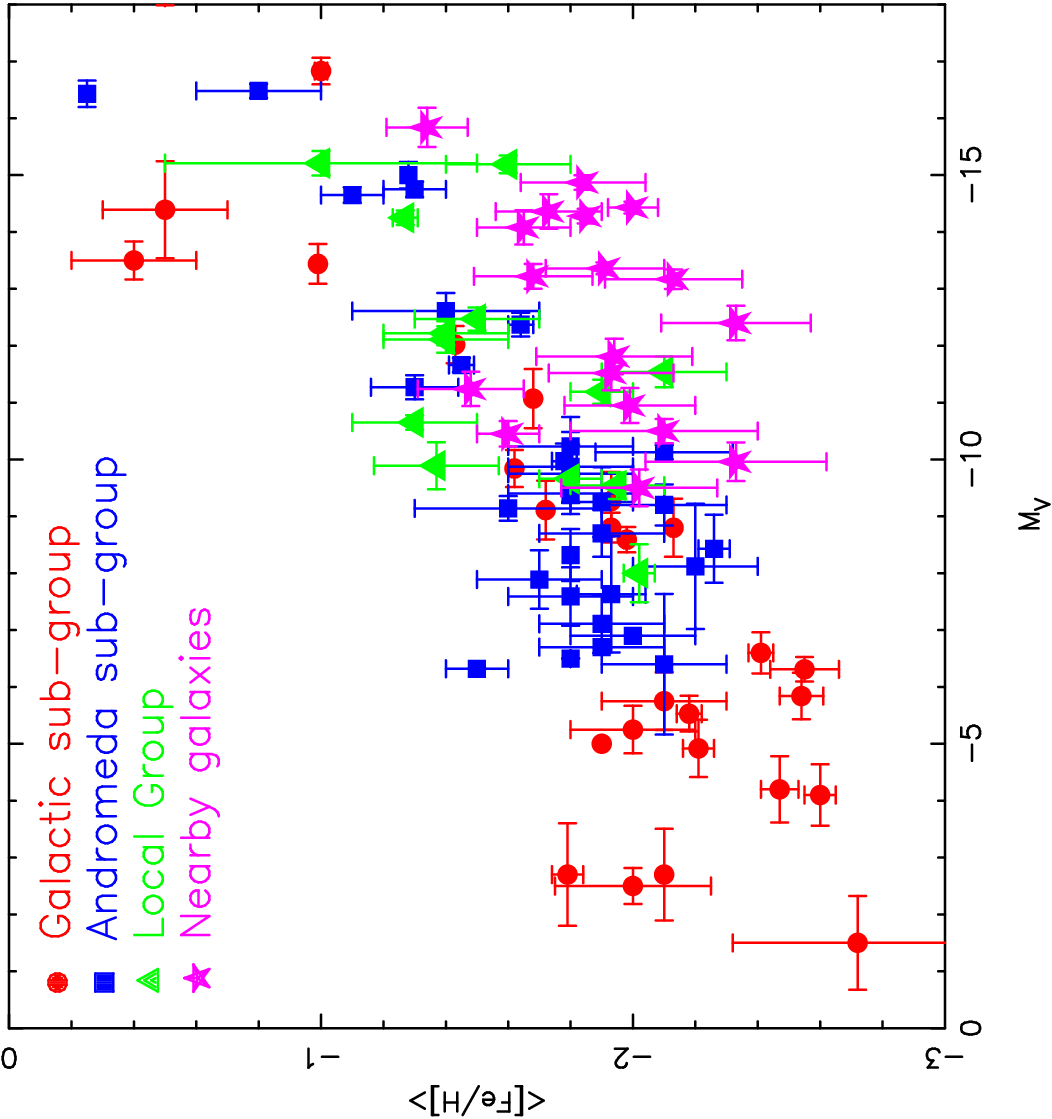}
    \includegraphics[angle=270, width=8cm]{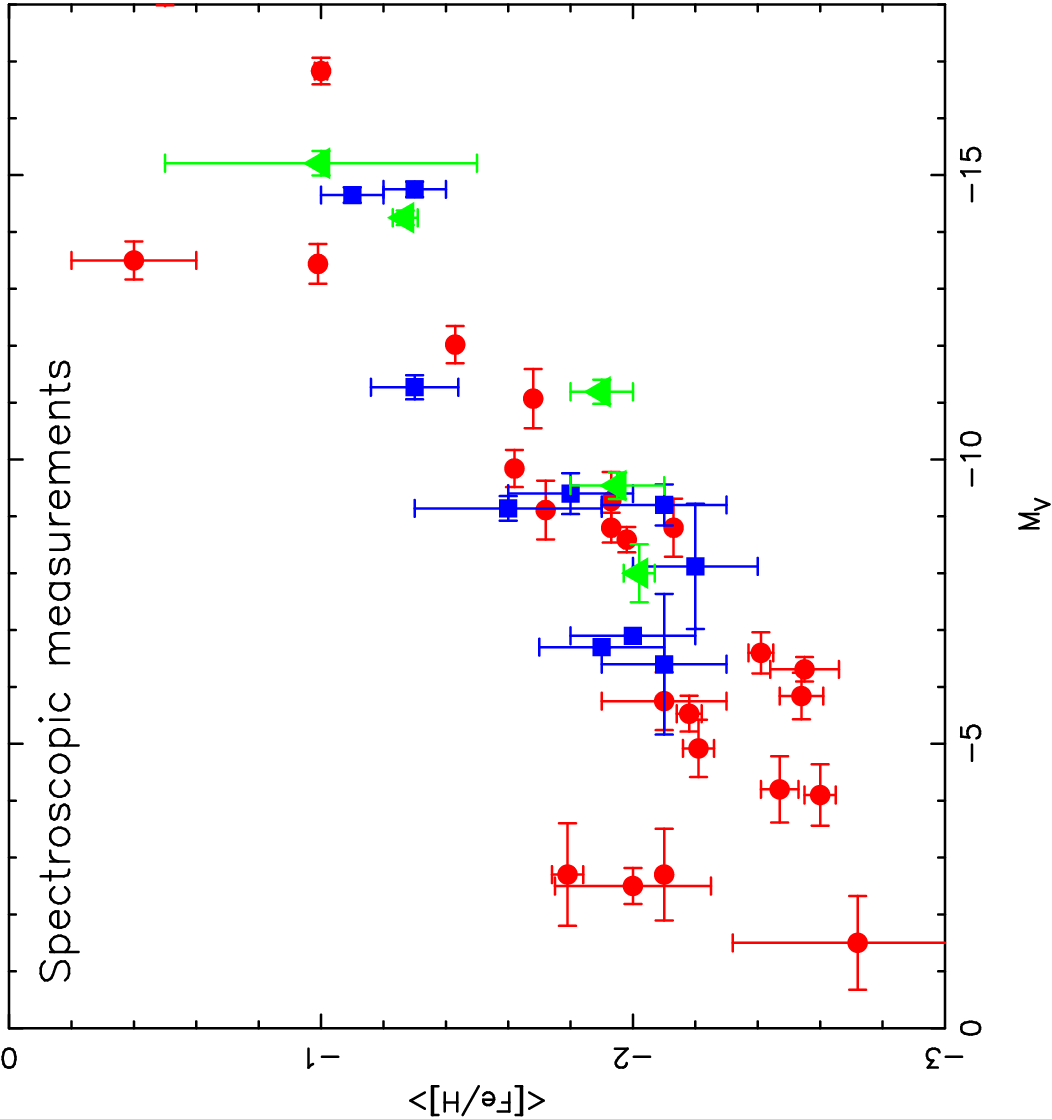}
    \caption{Absolute visual magnitude versus stellar metallicity
      measurement (as listed in Table 5), for all galaxies in the
      sample for which this information is available.  Color coding is
      the same as in previous figures. The left panel shows all
      measurements regardless of the techniques used to estimate
      metallicity, whereas the right panel only shows those
      measurements derived from spectroscopic observations of resolved
      stars (typically giants).}
  \end{center}
\end{figure*}


\clearpage


\begin{deluxetable}{llccrrll}
\rotate
\tabletypesize{\tiny}
\tablecaption{Basic information}
\tablewidth{625pt}
\tablehead{
\colhead{(1)} & \colhead{(2)} & \colhead{(3)} & \colhead{(4)} & \colhead{(5)} & \colhead{(6)} & \colhead{(7)} & \colhead{(8)}\\
\colhead{Galaxy} & \colhead{Other names} & \colhead{} & \colhead{} & \colhead{R.A.} & \colhead{Dec.} & \colhead{Original Publication} & \colhead{Comments}\\
\colhead{} & \colhead{} & \colhead{} & \colhead{} & \colhead{J2000} & \colhead{J2000} & \colhead{} & \colhead{}
}

\startdata

\multicolumn{6}{l}{{\it The Milky Way sub-group (in order of distance from the Milky Way)}}\\
The Galaxy            & The Milky Way            & G &S(B)bc& 17h45m40.0s & -29d00m28s & --- & Position refers to Sgr A$ ^*$ \\
Canis Major           &                          & G & ???? & 07h12m35.0s & -27d40m00s &\cite{martin2004a} & MW disk substructure?\\
Sagittarius dSph      &                          & G & dSph & 18h55m19.5s & -30d32m43s &\cite{ibata1994} & Tidally disrupting\\
Segue (I)             &                          & G & dSph & 10h07m04.0s & +16d04m55s &\cite{belokurov2007}\\
Ursa Major II         &                          & G & dSph & 08h51m30.0s & +63d07m48s &\cite{zucker2006a}\\
Bootes II             &                          & G & dSph & 13h58m00.0s & +12d51m00s &\cite{walsh2007}\\
Segue II              &                          & G & dSph & 02h19m16.0s & +20d10m31s &\cite{belokurov2009}\\
\enddata

\tablenotetext{a}{Originally thought to be a globular cluster. \cite{canterna1977} subsequently  identified its galactic nature.}
\tablenotetext{b}{This author first suggested that this object is a member of the Local Group, but it had previously been cataloged by \cite{corwin1985}.}
\tablenotetext{c}{These authors  calculate a new distance to UGC 4879 and show it is on the periphery of the Local Group. While included in the ``Atlas and Catalog of Interacting Galaxies'' by \cite{vorontsovvelyaminov1959}, earlier references to this object have not been found.}
\tablenotetext{d}{\cite{holmberg1958} credits \cite{wilson1955} with discovery, but this object is not listed in this manuscript.}
\tablenotetext{e}{These authors rediscovered this galaxy and showed that it was in the periphery of the Local Group, but it had earlier been catalogued by \cite{corwin1985}, \cite{feitzinger1985} and \cite{arp1987}.}
\tablenotetext{f}{HIZSS 3 resolved into two  sources by \cite{begum2005}. Position corresponds to the ``HIZSS3 system''.}
\tablenotetext{g}{HIZSS 3 resolved into two  sources by \cite{begum2005}. Position corresponds to the ``HIZSS3 system''.}
\tablenotetext{h}{See also \cite{karachentseva1999}.}
\tablenotetext{i}{Discovered on inspection of photographic plates from a  survey discussed in \cite{shane1947}. C.D. Shane speculated that some of the nebulae that were visible could be Local Group dwarf galaxies.}
\tablenotetext{j}{The original coordinates of NGC 4167 are coincident with the coordinates of NGC 4163, but \cite{sulentic1973} note that NGC 4167 is ``non-existent.''}

\end{deluxetable}


\clearpage
\begin{deluxetable}{lrrrrrrrrrcccl}
\tablewidth{625pt}

\tabletypesize{\scriptsize}
\rotate
\tablecaption{Positions and velocities}

\tablehead{
\colhead{(1)} & \colhead{(2)} & \colhead{(3)} & \colhead{(4)} & \multicolumn{2}{c}{(5)} & \multicolumn{2}{c}{(6)} & \multicolumn{2}{c}{(7)} & \colhead{(8)} & \colhead{(9)} & \colhead{(10)} & \colhead{(11)}\\
\colhead{Galaxy} & \colhead{l} & \colhead{b} & \colhead{$E(B-V)$} & \multicolumn{2}{c}{$(m-M)_o$} & \multicolumn{2}{c}{$D_\odot$} & \multicolumn{2}{c}{$v_\odot$} & \colhead{[$D_{MW}$,$v_{MW}$]} & \colhead{[$D_{M31}$,$v_{M31}$]} & \colhead{[$D_{LG}$,$v_{LG}$]} & \colhead{References\tablenotemark{a}}\\
\colhead{} & \colhead{} & \colhead{} & \colhead{} & \multicolumn{2}{c}{} & \multicolumn{2}{c}{kpc} & \multicolumn{2}{c}{km\,s$^{-1}$} & \colhead{kpc, km\,s$^{-1}$} & \colhead{kpc, km\,s$^{-1}$} & \colhead{kpc, km\,s$^{-1}$} & \colhead{}\\
}
\startdata
The Galaxy            &\\
Canis Major\tablenotemark{b} & 240.0 &  -8.0 & 0.264 & 14.29 & 0.30 &    7 &   1 &   87.0\tablenotemark{c} & 4.0 & [  13, -117] & [ 786, -185] & [ 394, -172] & (1) (2)\\
Sagittarius dSph      &   5.6 & -14.2 & 0.153 & 17.10 & 0.15 &   26 &   2 &  140.0 & 2.0 & [  18,  169] & [ 791,  149] & [ 400,  159] & (3) (4)\\
Segue (I)             & 220.5 & +50.4 & 0.031 & 16.80 & 0.20 &   23 &   2 &  208.5 & 0.9 & [  28,  113] & [ 792,   56] & [ 401,   69] & (5) (6)\\
Ursa Major II         & 152.5 & +37.4 & 0.094 & 17.50 & 0.30 &   32 &   4 & -116.5 & 1.9 & [  38,  -33] & [ 771,   11] & [ 380,   -3] & (7) (8)\\
Bootes II             & 353.7 & +68.9 & 0.031 & 18.10 & 0.06 &   42 &   1 & -117.0 & 5.2 & [  40, -116] & [ 807, -174] & [ 416, -156] & (9) (10)\\
Segue II              & 149.4 & -38.1 & 0.185 & 17.70 & 0.10 &   35 &   2 &  -39.2 & 2.5 & [  41,   43] & [ 753,  141] & [ 361,  112] & (11)\\
Willman 1             & 158.6 & +56.8 & 0.014 & 17.90 & 0.40 &   38 &   7 &  -12.3 & 2.5 & [  43,   35] & [ 780,   44] & [ 390,   41] & (12) (13)\\
\enddata
\end{deluxetable}

\clearpage


\begin{deluxetable}{lrrrrrrrrrrrrrrrll}
\tabletypesize{\tiny}
\rotate

\tablecaption{Luminosity and structural parameters}

\tablewidth{625pt}

\tablehead{
\colhead{(1)} &  \multicolumn{2}{c}{(2)} & \multicolumn{2}{c}{(3)} & \multicolumn{2}{c}{(4)} & \multicolumn{2}{c}{(5)} & \multicolumn{2}{c}{(6)} & \multicolumn{2}{c}{(7)} & \multicolumn{2}{c}{(8)} & \colhead{(9)} & \colhead{(10)} & \colhead{(11)}\\
\colhead{Galaxy} &  \multicolumn{2}{c}{$V$} & \multicolumn{2}{c}{$r_h$} & \multicolumn{2}{c}{$\mu_V$} & \multicolumn{2}{c}{$\theta$} & \multicolumn{2}{c}{$\epsilon$} & \multicolumn{2}{c}{$M_V$} & \multicolumn{2}{c}{$r_h$} & \colhead{$\mu_{eff}$} & \colhead{References\tablenotemark{a}} & \colhead{Comments}\\
\colhead{} & \colhead{} & \colhead{} & \multicolumn{2}{c}{'} & \multicolumn{2}{c}{$M/\square"$} & \multicolumn{2}{c}{$^\circ$} & \colhead{} & \colhead{} & \colhead{} & \colhead{} & \multicolumn{2}{c}{pc} & \colhead{$M/\square"$} & \colhead{} & \colhead{} \\
}

\startdata
The Galaxy\\
Canis Major            &  -0.1 &  0.8 &     --- &    --- &  24.0\tablenotemark{b} &  0.6 &  123:\tablenotemark{c} &     &   0.8: &       &  -14.4\tablenotemark{d} &  0.8 &   --- &  --- &  ---   & (1) (2)\\
Sagittarius dSph       &   3.6 &  0.3 &  342.00 &  12.00 &  25.2 &  0.3 &  102 &   2 &  0.64 &  0.02 &  -13.5 &  0.3 &  2587 &  219 &  26.0  & (3) (4)\\
Segue                  &  15.3 &  0.8 &    4.40 &$^{+1.2}_{-0.6}$&  27.6 &  1.0 &   85 &   8 &  0.48 &  0.13 &   -1.5 &  0.8 &    29 &$^{+8}_{-5}$&  28.7  & (5)\\ 
Ursa Major II          &  13.3 &  0.5 &   16.00 &   1.00 &  27.9 &  0.6 &   98 &   4 &  0.63 &  0.05 &   -4.2 &  0.6 &   149 &   21 &  29.1  & (5)\\
Bootes II              &  15.4 &  0.9 &    4.20 &   1.40 &  28.1 &  1.6 &  145 &  55 &  0.21 &  0.21 &   -2.7 &  0.9 &    51 &   17 &  29.1  & (5)\\
Segue II               &  15.2 &  0.3 &    3.40 &   0.20 &  27.4 &  0.4 &  182 &  17 &  0.15 &  0.10 &   -2.5 &  0.3 &    35 &    3 &  28.6  & (6)\\
Willman 1              &  15.2 &  0.7 &    2.30 &   0.40 &  26.1 &  0.9 &   77 &   5 &  0.47 &  0.08 &   -2.7 &  0.8 &    25 &    6 &  27.2  & (5)\\
\enddata
\end{deluxetable}


\clearpage

\begin{deluxetable}{lrrrrrrrrrrclp{2in}}
\tablewidth{625pt}
\setlength{\hoffset}{-0.5in}
\setlength{\textwidth}{4.5in}
\tabletypesize{\tiny}
\rotate
\tablecaption{Masses and kinematics}

\tablehead{
\colhead{(1)} & \colhead{(2)} & \multicolumn{2}{c}{(3)} &\multicolumn{2}{c}{(4)} & \colhead{(5)} & \multicolumn{2}{c}{(6)} & \multicolumn{2}{c}{(7)} & \colhead{(8)} & \colhead{(9)} & \colhead{(10)}\\
\colhead{Galaxy} & \colhead{$M_\star$} & \multicolumn{2}{c}{$\sigma^*$} & \multicolumn{2}{c}{$v_r^*$} & \colhead{$M_{HI}$} & \multicolumn{2}{c}{$\sigma^{HI}$} & \multicolumn{2}{c}{$v_r^{HI}$} & \colhead{$M_{dyn}(\le r_h)$} & \colhead{References\tablenotemark{a}} & \colhead{Comments}\\
\colhead{} &  \colhead{$10^6M_\odot$} & \multicolumn{2}{c}{km\,s$^{-1}$} & \multicolumn{2}{c}{km\,s$^{-1}$} & \colhead{$10^{6}M_\odot$} & \multicolumn{2}{c}{km\,s$^{-1}$} & \multicolumn{2}{c}{km\,s$^{-1}$} & \colhead{$10^{6}M_\odot$} & \colhead{} & \colhead{} \\
}

\startdata
The Galaxy\\          
Canis Major       &       49 & 20.0 &  3.0 &  --- & ---  &    --- & --- & --- & --- & --- &     --- & (1) & HI measurements complex due to location!\\
Sagittarius dSph  &       21 & 11.4 &  0.7 &  N/A & N/A  &    N/A\tablenotemark{b}  & N/A & N/A & N/A & N/A &     190 & (2) (3) (4) (5) & \\
Segue (I)         &  0.00034 &  3.9\tablenotemark{c} &  0.8 &  N/A & N/A  &    N/A & N/A & N/A & N/A & N/A &    0.26 & (6) (5) & Stellar velocity gradient constrained to $<5$\,km\,s$^{-1}$ at 90\% confidence\\
Ursa Major II     &   0.0041 &  6.7 &  1.4 &  N/A & N/A  &    N/A & N/A & N/A & N/A & N/A &     3.9 & (7) (5) &Velocity difference of $8.4 \pm 1.4$km\,s$^{-1}$ between stars in the east and west\\
Bootes II         &   0.0010 & 10.5 &  7.4 &  --- & ---  &    N/A & N/A & N/A & N/A & N/A &     3.3 & (8) (5) & \\		
Segue II          &  0.00086 &  3.4 & $^{+2.5}_{-1.2}$&  --- & ---  &    --- & --- & --- & --- & --- &    0.23 & (9) \\
Willman 1         &   0.0010 &  4.3 & $^{+2.3}_{-1.3}$ &  --- & ---  &    N/A & N/A & N/A & N/A & N/A &    0.27 & (10) (5) & \\
\enddata
\end{deluxetable}

\clearpage


\begin{deluxetable}{lrrllp{1.5in}}
\tablewidth{450pt}

\tabletypesize{\scriptsize}

\tablecaption{Mean stellar metallicities}

\tablehead{
\colhead{(1)} & \multicolumn{2}{c}{(2)} & \colhead{(3)} & \colhead{(4)} & \colhead{(5)}\\
\colhead{Galaxy} & \multicolumn{2}{c}{$<[Fe/H]>$} & \colhead{Technique} & \colhead{References} & \colhead{Comments}\\
}

\startdata
The Galaxy         &&&&&\\
Canis Major        & -0.5  & 0.2  & CMD modeling & \cite{bellazzini2004c} & Based primarily on MS stars\\
Sagittarius dSph   & -0.4  & 0.2  & HRS & \cite{chou2007}\tablenotemark{a}\\
Segue (I)          & -2.72 & 0.40 & MRS & \cite{norris2010} & See also
\cite{simon2011}\\
Ursa Major II      & -2.47 & 0.06 & MRS       & \cite{kirby2008b, kirby2011}\\
Bootes II          & -1.79 & 0.05 & MRS (CaT) & \cite{koch2009}\\
Segue II           & -2.00 & 0.25 & HRS & \cite{belokurov2009}\\
Willman 1          & -2.1: &      & MRS & \cite{willman2011} & Based on three stars\\
\enddata

\tablenotetext{a}{And references therein.}

\end{deluxetable}

\end{document}